\documentclass{article}

\usepackage{arxiv} %

\usepackage[utf8]{inputenc} %
\usepackage[T1]{fontenc}    %
\usepackage{array}          

\usepackage{amsfonts}       %
\usepackage{amsmath}        

\usepackage{booktabs}       
\usepackage{graphicx}       
\usepackage{nicefrac}       

\usepackage{caption}
\usepackage{natbib}         %
\usepackage{doi}            %

\usepackage{hyperref}       %
\hypersetup{hidelinks}      %

\usepackage{microtype}      %
\usepackage{url}            %
\usepackage{lipsum}         %

\usepackage[nottoc]{tocbibind} 

\usepackage[most]{tcolorbox}
\usepackage{enumitem}

\renewcommand{\headeright}{} %
\renewcommand{\undertitle}{} %
\renewcommand{\shorttitle}{} %

\title{What is an Intelligent System?}

\author{ {Martin Molina} \\
	Department of Artificial Intelligence\\
	Universidad Politécnica de Madrid - Spain\\
}

\date{August 2026}

\renewcommand{\headeright}{}
\renewcommand{\undertitle}{}
\renewcommand{\shorttitle}{}

\hypersetup{
pdftitle={What is an Intelligent System?},
pdfauthor={Martin Molina},
pdfkeywords={Artificial Intelligence, AI systems},
}

\begin{document}

\vspace*{\fill} 
\begin{center}
    \begin{minipage}{0.85\textwidth} 
        \maketitle
        \vspace{1cm}
        \begin{abstract}
            The term \textit{intelligent system} has emerged in the field of information technology as a category of computer systems derived from successful applications of artificial intelligence. This paper proposes a general description that identifies the main properties and types of components typically found in such systems. Adopting an integrative and pedagogical approach, this description provides a conceptual framework for systems engineering practitioners seeking a coherent vocabulary and organizational structure to approach the analysis and construction of intelligent systems. The paper presents examples of both classical and modern intelligent systems to illustrate the generality and applicability of the description.

        \end{abstract}
    \end{minipage}
\end{center}
\vspace*{\fill}
\thispagestyle{empty} 
\clearpage


\newpage
\pagenumbering{roman}
\setcounter{tocdepth}{3} 

\begingroup
\renewcommand{\baselinestretch}{0.1} 
\setlength{\parskip}{3pt}
\tableofcontents 
\endgroup

\setcounter{page}{1} 
\newpage
\pagenumbering{arabic} 

\section{Introduction} \label{ref:introduction}
The historical progress of humanity has been closely linked to the creation of increasingly powerful and sophisticated tools. In the age of the industrial revolution, a large number of tools were built as machines that automate tasks that require physical effort. In the digital age, computer systems are being created to automate tasks that require mental effort. The capabilities of these systems have been progressively increased to perform tasks that require more and more intelligence. This evolution has generated a type of tool that we call an intelligent system.

Intelligent systems help us perform specialized tasks in professional domains such as medical diagnosis (e.g., recognizing tumors in radiological images) or airport management (e.g., generating a new assignment of airport gates in the presence of an incident). They can also perform tedious tasks for us (e.g., autonomous car driving or house cleaning) or dangerous tasks such as the exploration of unknown areas (e.g., underwater exploration).

The development of such systems is now an engineering discipline of information technology that requires effective methods and tools. The precise characterization of what is meant by an intelligent system is non-trivial because it is based on concepts related to cognition, an area whose terminology and level of abstraction may differ according to the area of study (neuroscience, computer science, robotics, philosophy, cognitive psychology, etc.). Some of the terms used can even change with the proposal of new computational models of intelligence and new scientific findings related to our understanding of the mind. This often leads to misunderstandings between systems engineers because they interpret the same terms in different ways.

The main goal of this paper is to propose a general description of the properties and types of components commonly observed in intelligent systems, integrating previous conceptions. Its purpose is primarily pedagogical, aiming to provide students and systems engineering practitioners with a unified conceptual framework for organizing ideas that are sometimes dispersed across multiple documentary sources. The characterization presented here is based on well-established concepts from the literature and synthesizes them into a structure designed to facilitate the analysis, design, and communication of the practical applications of these systems.

The remainder of the paper is organized as follows. Section \ref{ref:approaches} presents a review of the literature on what an intelligent system is meant to be in artificial intelligence and concludes by describing the criteria followed in this work to characterize such systems from multiple complementary views. These views are developed in detail from Section \ref{ref:operational_context} to Section \ref{ref:architectural_form}. Finally, Section \ref{ref:examples} describes examples of intelligent systems using the presented characterization, covering both classical and modern systems.

\section{Criteria used to define an intelligent system} \label{ref:approaches}
The meaning of intelligent system varies depending on the context. This document focuses on artificial systems within an engineering context, distinct from that of other disciplines (such as neuroscience, psychology, sociology, biology, or philosophy), where intelligent behavior is analyzed as a natural phenomenon.

In this context, the term \textit{system} is understood as an artifact designed and built from a set of interrelated components to achieve a specific purpose. For example, in software engineering, it can be understood as a collection of components organized to accomplish a specific function or set of functions\footnote{ IEEE Standard Glossary of Software Engineering Terminology: \url{https:/standards.ieee.org/ieee/610.12/855}}.

\begin{figure}[htb!]
\begin{center}
\includegraphics[width=0.65\textwidth]{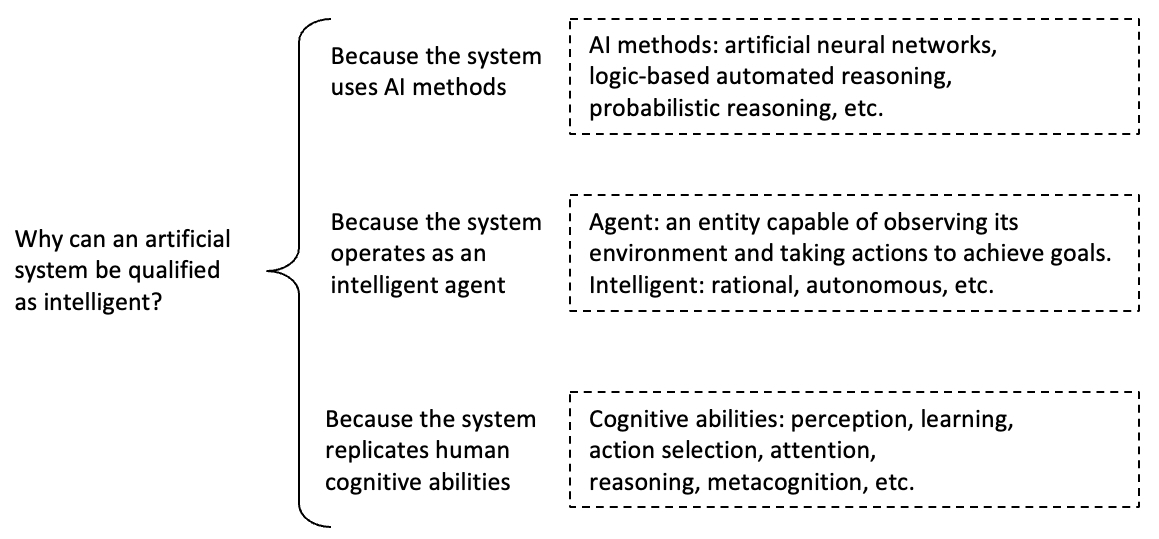}
\caption{Commonly used reasons for deciding whether a particular artificial system can be qualified as intelligent.}
\label{fig:approaches_to_define}
\end{center}
\end{figure}

To decide whether an artificial system can be qualified as \textit{intelligent}, three criteria with varying degrees of acceptance can be distinguished (Figure \ref{fig:approaches_to_define}): (1) the system uses artificial intelligence methods, (2) the system operates as an intelligent agent, or (3) the system replicates cognitive capabilities observed in humans. These criteria are detailed in the following sections, concluding with a description of the integrated approach proposed in this paper.

\subsection{Systems that use artificial intelligence methods} 

An artificial system could be considered intelligent simply because it uses artificial intelligence methods. This criterion is based on the nature of the elements that compose the system. For example, a system that uses a neural network to classify an image and a system that uses logic programming to diagnose a given malfunction could be considered intelligent systems.

This is a direct way of classifying a system as intelligent, and it is used, for example, in the naming of scientific journals and conferences that publish advances in artificial intelligence methods\footnote{Some of these journals are \textit{IEEE Intelligent Systems} (IEEE Computer Society), \textit{Intelligent Systems with Applications} (Elsevier publisher), and \textit{International Journal of Intelligent Systems} (Wiley publisher).}. It is also used in the naming of courses in university programs describing artificial intelligence methods. In recent years, the use of terms like \textit{AI system} or \textit{AI-based system}\footnote{For example, the standard ISO/IEC TR 29119-11:2020 defines an AI-based system as a system including one or more components implementing AI.} has also become popular, which have a similar meaning.

The main limitation of this definition is that it is circular. It states that a system is intelligent because it uses AI methods, but these methods are called AI because they were designed to generate intelligent behaviors. Furthermore, classifying a system as intelligent based on its internal technical components (such as neural networks or logic) ignores its performance, giving a label of intelligence for using a specific tool rather than for successfully solving problems. The next sections show other ways of determining when an artificial system may be called intelligent that are not based on the nature of its components but on the way the system operates.

\subsection{Systems that operate like intelligent agents}

In the AI community, the notion of an agent has been commonly used to define and describe the characteristics of intelligent systems. The term \textit{agent} can be understood as an entity capable of observing its environment and taking actions to achieve its goals\footnote{According to its etymological origin, the word agent means \textit{someone who acts} (e.g., the Latin word \textit{agens} is the present active participle of \textit{\href{https://en.wiktionary.org/wiki/ago\#Latin}{agere}} meaning to do or perform).}. The idea of an agent has also been described in AI as a system that acts to perform tasks on behalf of its user\footnote{This approach has been studied in economics (and other disciplines such as political science, law, etc.) as the \textit{principal-agent} model of the theory of agency \citep{Ross-1973}  \citep{Eisenhardt-1989,Shapiro-2005}, in which one party, called \textit{the agent}, acts on behalf of another party called the \textit{principal}} \citep{Nwana-1996,Gabriel-2024}. In recent years, the concept of an agent has been approached in a more flexible way, considering that there are different degrees to which a system can be \textit{agentic} \citep{Chan-2023,Shavit-2023}.

According to some authors, an essential characteristic of an agent to be qualified as intelligent is that it \textit{acts rationally} \citep{Newell-1982,Russell-2014}. An agent acts rationally when its decisions, based on available information about the environment (usually partial and uncertain) and subject to its resource constraints, are oriented toward optimizing the degree of success in achieving its goals. For example, in the case of a self-driving car that selects the route to follow, the degree of success can be quantified by the time it takes to reach the destination (besides other factors such as safety, comfort, fuel consumption, and toll costs).

This view of intelligence as the ability to achieve goals is present, with certain variations, in definitions of intelligence proposed by different authors. For example, Albus understands intelligence as the ability of a system to select actions that increase the probability of achieving goals in an uncertain environment \citep{Albus-1991}. Goertzel considers intelligence to be the achievement of complex goals in complex environments \citep{Goertzel-2006} and Kurzweil says that intelligence is the ability to use optimally limited resources to achieve goals \citep{Kurzweil-2000}. 

One practical advantage of this interpretation is that it provides an objective method for evaluating system intelligence based on observed external behavior, which avoids dealing with other not well-understood aspects of the human mind\footnote{This approach is followed by the benchmarks that have become widely used to measure technical progress in AI, which quantify a system's performance on specific tasks, often comparing it to human performance on those tasks.}. However, this approach does not provide a general classification criterion, since the measure of success depends on the type of task performed by each system\footnote{The results of multiple benchmarks can be aggregated into a single composite metric (such as the Artificial Analysis Intelligence Index: \url{https://artificialanalysis.ai/methodology/intelligence-benchmarking}). However, the choice of which tasks to include and how to weight them is a design decision on which there is no consensus.}. In addition, the degree of intelligence can be determined not only by how well an agent performs a task but also by the diversity of tasks that the agent is capable of performing, which also presents practical difficulties for its evaluation \citep{Legg-2007}. 

Beyond rational action, other types of capabilities have been considered to qualify an agent as intelligent, although there is no agreement among different authors regarding which of these are essential or what relative weight should be assigned to each. Other capabilities considered by various authors include autonomy, reactivity, proactivity, and social ability to interact with other agents \citep{Wooldridge-1999,Wooldridge-1995,Franklin-1996}. 

\subsection{Systems that replicate human cognitive abilities}

Another way of looking at the intelligence of a system is by comparison with human intelligence, which is usually understood as a complex ability resulting from the combination of multiple cognitive abilities (e.g., perception, reasoning and learning) \citep{Gottfredson-1997,HayesRoth-1995,Anastasi-1992}. 

These abilities are formally studied in cognitive science, an area that integrates experimental techniques from disciplines such as neuroscience and psychology, together with computer models from artificial intelligence.  Multiple theories of mind have been proposed in the form of cognitive architectures \citep{Newell-1990} that are formulated with mechanisms described as information processes \citep{Laird-2017,Kotseruba-2020}. The usual cognitive abilities considered in these architectures are perception, attention, action selection, memory, learning, reasoning, and metacognition \citep{Vernon-2022}. Other classifications use more specific categories of cognitive abilities, such as recognition, categorization, decision making, perception, situation assessment, prediction, monitoring, planning, executing actions, communication, remembering, belief maintenance, reflection, or learning \citep{Langley-2009}. 

For developers of artificial systems, these proposals serve as a useful reference for designing generic computational mechanisms that support intelligent behavior. However, it is not easy to directly use a complete cognitive architecture to build a final system, as the proposed architectures are more oriented toward explaining the mechanisms of the mind than serving as a guide for constructing systems that operate efficiently. Furthermore, there is no consensus on which of these proposals should be used as a reference for characterizing the capabilities of an intelligent system.

\begin{table}[h]
\centering
\footnotesize
\renewcommand{\arraystretch}{1.5} 
\setlength{\tabcolsep}{5pt}      
\begin{tabular}{lp{5cm}p{6.5cm}} 
\toprule
\textbf{View} & \textbf{Scope}& \textbf{Description}\\ 
\midrule
Operational context & 
This view identifies the primary function of the system, describing how it interacts with external entities.& 
An intelligent system operates as an agent that helps the user perform tasks in a complex work environment. The system can modify the environment's state or recommend how to do it to achieve user-defined objectives.\\ \addlinespace
Behavioral properties & 
This view describes the properties of the system's behavior from an external perspective.& 
An intelligent system can behave rationally, be autonomous, learn (improving its performance through experience), and explain its conclusions.\\ \addlinespace
Basic functions & 
This view describes what the system does from an internal functional perspective.& 
To operate, an intelligent system can perceive its environment, evaluate the situation, decide on actions to take, and execute actions in a controlled manner.\\ \addlinespace
Architectural form & 
The architectural form of the system describes the characteristics of the components used in its construction. & 
The architecture of an intelligent system includes model-based components with general inference processes. Models can be generated from datasets through machine learning. \\ 
\bottomrule
\end{tabular}
\caption{Summary of the characterization of an intelligent system using four complementary views.}
\label{tab:definition_of_intelligent_system}
\end{table}

\subsection{Characterization proposed in this paper}

The previous sections show that the revised criteria are not fully satisfactory for establishing a general definition for classifying a system as intelligent due to problems such as circularity, a lack of consensus, and difficulties in their practical application. Therefore, rather than proposing a classificatory definition of what an intelligent system is, this document adopts a descriptive approach aimed at identifying the general properties and types of components that typically characterize these systems. The goal is to provide a practically useful reference for systems engineering professionals that facilitates both the analysis of existing systems and the development of new intelligent systems.

According to this, this document proposes a characterization divided into four separate views that address different aspects of the system, following a common practice in systems engineering for managing complexity during the analysis and design phases. This separation makes it possible to look at each aspect independently instead of mixing up issues that are often confused with one another. For example, the separation helps distinguish how the system interacts with its environment from the properties that make its behavior perceived as intelligent or from the internal mechanisms that make such behavior possible. Table \ref{tab:definition_of_intelligent_system} summarizes the four views considered.

The following sections of this article develop the four views of this characterization, providing a description that explains each of them in greater detail. Section \ref{ref:examples} presents examples of intelligent systems that are analyzed using the characterization proposed in this document.

\section{Operational context of an intelligent system} \label{ref:operational_context}
The operational context of an intelligent system identifies the classes of interactions that the system performs with external elements. To describe this context, the intelligent system is conceived as an agent that is part of a global system. Within the global system, two other subsystems are identified: the \textit{user} and the work \textit{environment} on which the intelligent system operates. In general, it is assumed that: (1) the user (human or machine) defines the tasks to be performed by the intelligent system, and (2) the intelligent system acts on its environment to perform these tasks. 

\begin{figure}[htb!]
\begin{center}
\includegraphics[width=0.70\textwidth]{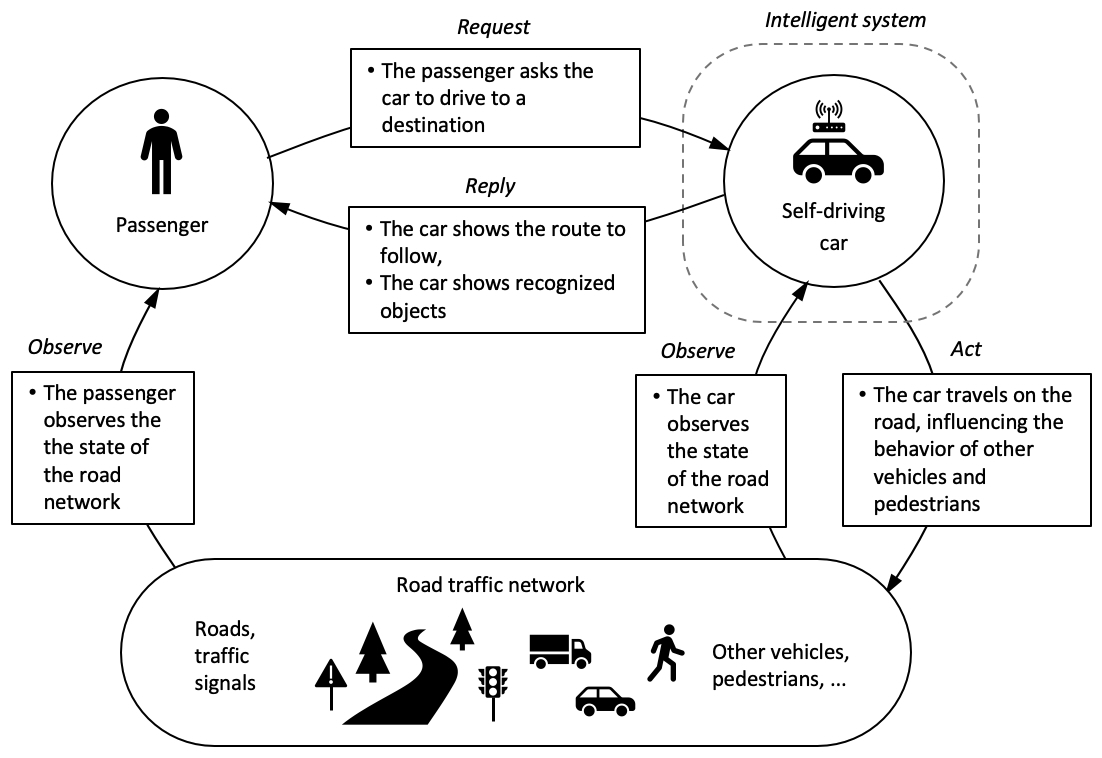}
\caption{Diagram used to describe the operational context of an intelligent system, applied to the case of an autonomous car. In this diagram, each node (circle or rounded box) represents a system and each arc represents an relationship of influence from one system to another. The rectangle over each arc explains what each system does to generate that influence.}
\label{fig:example_self_driving_car}
\end{center}
\end{figure}

Figure \ref{fig:example_self_driving_car} illustrates the operational context for the case of an autonomous car. The passenger acts as the user, asking the car to go to a certain destination, and the vehicle moves through the road network, which is considered its environment (with roads, traffic lights, pedestrians, other vehicles, etc.).

The following sections detail the characteristics of certain typical forms of interaction between these subsystems, analyzing the factors that increase their complexity and the technical terminology used in their description. 

\subsection{Interaction with the environment}

As shown in the previous example, a self-driving car interacts with the system that represents the road traffic network where it circulates. The car observes the state of the road network by means of cameras and other sensors (GPS sensors, LiDAR sensors, ultrasonic sensors, etc.) and acts on the environment, traveling on the road and influencing the behavior of other vehicles and pedestrians.

\begin{figure}[htb!]
\begin{center}
\includegraphics[width=0.50\textwidth]{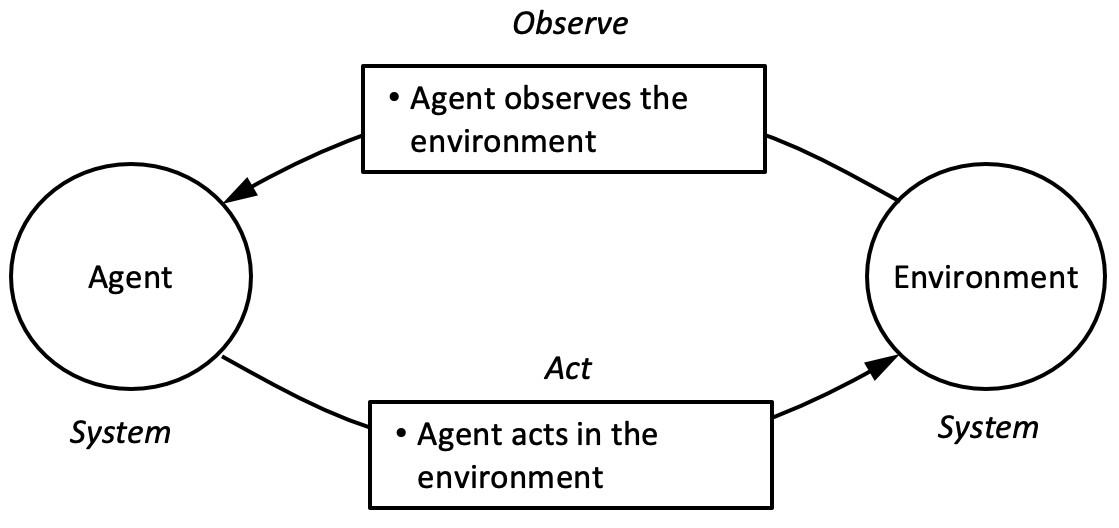}
\caption{Form of interaction between systems according to the agent-environment pattern.}
\label{fig:pattern_agent_environment}
\end{center}
\end{figure}

This form of interaction can be understood in a general way according to the pattern in Figure \ref{fig:pattern_agent_environment}. This pattern has been described in the artificial intelligence literature to show the basic way in which an agent operates \citep{Wooldridge-1999,Russell-2014}. In this pattern, there are two systems with the roles of agent and environment. The environment represents the set of surrounding objects, both static and dynamic, with which the agent can interact to perform its work. The interaction between the systems is bidirectional:
\begin{itemize}
\item \textit{Observe}: The agent can obtain data (\textit{observations}) about the state of objects in the environment (e.g., by means of sensors). 
\item \textit{Act}: The agent can execute \textit{actions} that change the state of objects in the environment (e.g., with the help of actuators). 
\end{itemize}

It is important to highlight that, in the context of intelligent systems, the working environment is usually complex for the agent. One of the factors influencing complexity is that the environment is \textit{dynamic}, that is, the state of the environment changes continuously, independently of the actions the agent may take, and these changes influence the agent's decisions on how to act. In this case, the agent must periodically observe the environment with a frequency high enough to be able to perform the task for which it is designed. An agent is said to be \textit{situated} in the environment when it operates in close interaction with the environment in a continuous sequence of observations and actions.

Another factor that influences the complexity of this interaction is the agent's resource limitations. Observation mechanisms (e.g., sensors) may provide incomplete or noisy information. The computers used may exhibit limitations in their computational capacity, either due to insufficient processing power or available memory. Furthermore, actuator limitations may prevent actions from being performed with total precision and dexterity. In these cases, the system must be capable of operating with a partial and uncertain observation of its environment and may need to execute a greater number of actions to successfully complete the tasks requested by the user.

\subsection{Interaction with the user}

In the self-driving car example, the user of the vehicle is the passenger who wishes to travel to a certain destination. The passenger indicates to the vehicle the desired destination and, in response, the vehicle communicates different types of information to the passenger, such as the planned route, the interpretation of data from the environment (showing recognized objects in the environment, such as pedestrians, traffic signs, etc.), a warning message when the destination is reached, etc.

\begin{figure}[htb!]
\begin{center}
\includegraphics[width=0.50\textwidth]{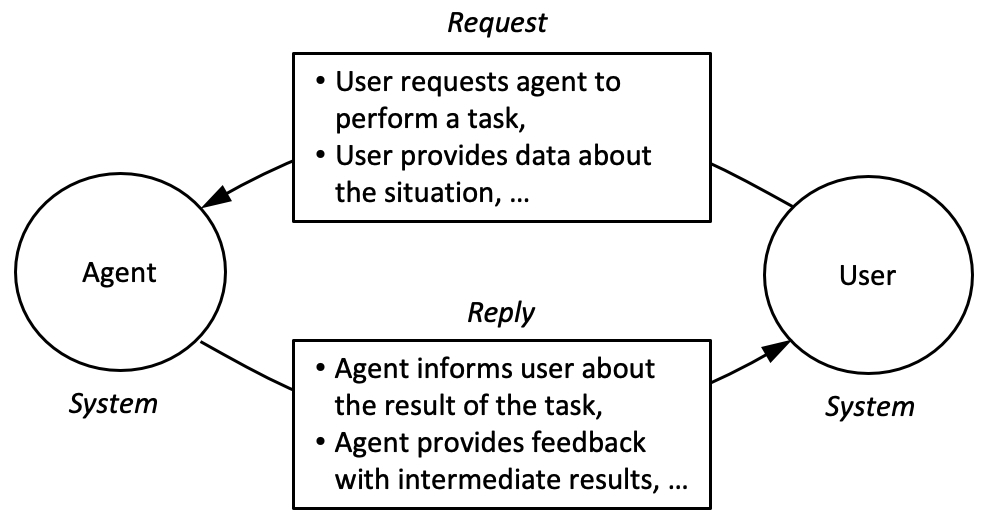}
\caption{Form of interaction between systems according to the agent-user pattern.}
\label{fig:pattern_agent_user}
\end{center}
\end{figure}

This form of interaction can be represented by the agent-user pattern (Figure \ref{fig:pattern_agent_user}) with the following types of influence relations:
\begin{itemize}
\item \textit{Request}. The user asks the agent to perform a \textit{task}. The task expresses an abstract action (for example, transporting a passenger, diagnosing a patient, or maintaining the temperature of a room) which may include a goal state (for example, the destination location of a vehicle or the desired heating temperature). This request is understood in a broad sense, since the user can also provide data about the situation in the environment that is not directly observable by the agent.
\item \textit{Respond}. The agent generates a response as a result of the performed task. The agent can also generate information that helps the user monitor the progress of the work or understand the decisions the agent makes (providing, for example, information on intermediate results during task execution).
\end{itemize}

The agent-user interaction pattern identifies a way of operating the intelligent system according to a superior-subordinate scheme in which the user delegates part of the work to be performed to the intelligent system. This form of interaction has been described in the AI assistant model \citep{Gabriel-2024,Nwana-1996} or in the principal-agent model of agency theory \citep{Eisenhardt-1989,Shapiro-2005}.

Note that the user could be viewed as part of the agent's environment and, therefore, the agent-user pattern could be considered within the more general agent-environment pattern. However, detailing the agent-user interaction separately is useful to characterize the properties of intelligent systems in a more precise way. For example, this form of interaction provides an adequate context to describe that the agent and the user use a common symbolic language to refer to objects in the environment.

In general, the user formulates the task at an abstract level with coarse granularity (for example, ``take me to the airport''), while the elementary actions that the agent must execute in the environment operate at a more detailed level with fine granularity (for example, acceleration, braking, and steering actions). Between both levels, there is an \textit{abstraction gap} such that the wider the gap, the greater the translation effort the user delegates to the agent to decompose the requested task into executable actions. In a simple system such as a thermostat, the abstraction gap is narrower, given that the task requested by the user (e.g., ``maintain temperature at 21 °C'') is immediately translated into actions (it is enough to compare the measured temperature with the target threshold to determine the heater's on or off action).

Another important factor that influences the complexity of the interaction between the user and the agent is the type of language used in communication. Open natural language, for example, presents significant technical challenges for machines to process reliably and efficiently. Communication can also be complex if it combines multiple modes (voice, gestures, written messages, etc.) that must be integrated to achieve a correct interpretation of the user's messages.

\subsection{Recursive application of interaction patterns}

The agent-environment and agent-user patterns can be applied recursively to model complex intelligent systems. The following examples illustrate this idea:

\begin{itemize}
    \item \textit{Example} 1. An engineer describes an autonomous driving vehicle as a system composed of two subsystems: (1) agent \textit{A}, which acts as an autopilot responsible for planning the global route and supervising its execution, and (2) agent \textit{B}, which directly controls the physical driving mechanisms. In this scenario, the user of agent \textit{A} is the passenger of the vehicle, who indicates the trip destination to the system. In turn, agent \textit{A} operates as a user of agent \textit{B} since it requests that agent B perform abstract tasks expressed as motion references (such as desired trajectories and speeds), which agent \textit{B} ultimately translates into physical actions of acceleration, braking, and steering.
    \item \textit{Example} 2. The traffic control system of an urban intersection is modeled using two subsystems. Agent \textit{A} is the operational traffic light controller. It perceives the local traffic flow through sensors and decides, cycle by cycle, when to change phases within pre-established time margins. Agent \textit{B} observes the performance of agent \textit{A} over several cycles. If the observed performance of agent \textit{A} is not satisfactory (for example, because the red phase generates long queues), agent \textit{B} reconfigures the time margins of agent \textit{A}, expanding or reducing the minimum or maximum duration of each phase to improve future performance. In this case, the environment of agent \textit{A} is the traffic network, while the environment of agent \textit{B} is agent \textit{A}.
\end{itemize}

With the subsystem organization of the first example, the task that the human user requests from the vehicle (traveling to a certain destination) is translated among several agents into the specific final actions, each managing a portion of the total translation. This decomposition helps to design and validate the global system in a modular way.

The interactions shown in the second example are present in software systems in the form of reflective architectures, in which a meta-level maintains a representation of the base system and acts upon it to modify its behavior. This organization is related to the notion of metacognition, understood as the capacity that people have to reflect on their own way of thinking or acting in order to correct it.

The practical utility of the organization in the second example is that agent \textit{A} can remain simple, with a fixed logic that works well under certain conditions. If conditions change in a sustained manner, agent \textit{B} can easily modify that logic without having to redesign it from scratch. In this way, the system has the flexibility to adapt to diverse conditions while maintaining a simple control logic.

These examples show complex intelligent systems with subsystems whose operational context, whether their user or their environment, is constituted by other subsystems also understood as agents\footnote{In the field of \textit{multi-agent systems}, other more specific forms of interaction between agents are studied, based, for example, on collaboration or competition between agents, which are not described here as they exceed the scope of this document.}. It is important to note that the identified subsystems should be understood as configurations made by the developer to facilitate analysis, design, and implementation. The developer decides the boundaries separating each intelligent system from the subsystems with which it interacts, including or excluding specific interactions as a strategy to manage complexity.

\section{Intelligent behavior} \label{ref:behavioral_properties}
The previous section characterizes the primary function of the intelligent system, describing how it interacts with its environment and with the user. This section describes properties of the system's behavior that are externally observable and are commonly associated with intelligent behavior.

Four properties are examined here: rational action, learning, autonomy, and explainability. These properties frequently appear in the AI literature and are particularly useful for the characterization purpose adopted in this document. Taken together, they respectively describe an intelligent system as a system that acts effectively under information and resource constraints, improves its performance with experience, operates without continuous user intervention, and can justify the results it produces.

These properties are not proposed as a formal or exhaustive definition of intelligence. They are, rather, dimensions that allow for the objective and gradual comparison of partial aspects of intelligent behavior between systems. A system that selects more effective actions, learns more quickly, requires less user intervention, or provides better explanations tends to be perceived as more intelligent than another that does not

It is important to note that each of these properties admits an analysis from two complementary perspectives: (1) the perspective of an external evaluator, who can analyze the system’s observable behavior, and (2) the perspective of the developer, who can examine the internal mechanisms on which such behavior is based. The following subsections detail these four properties, analyzing them from an external perspective. The analysis of internal mechanisms is addressed in subsequent sections. At the end of this section, a subsection is included with other properties that the literature has also related to intelligent behavior.

\subsection{Rational acting}

As previously introduced, some authors consider rational acting to be an essential property of intelligent behavior \citep{Newell-1982,Russell-2014}. According to the operational context defined in this work, we can use the following definition:
\begin{quote}
\textbf{Acting rationally}. An agent acts rationally when performing a task $T$ if, given the information available from the environment (which is usually partial and uncertain) and its capacity to process that information, it selects the actions that maximize a performance metric $P$ that quantifies its success in executing task $T$.
\end{quote}
The performance metric $P$ depends on the class $C$ to which the task $T \in C$ executed by the agent belongs. For example, in the case of an autonomous car, whose task class is passenger transportation, the metric could be travel time (expressed as a negative value, since the goal is to minimize time). In the case of an investment recommendation system, the metric could be the financial profit obtained. The performance measure should be understood as an external criterion for success defined for the task class $C$. Its selection is part of the system specification and may be a combination of multiple factors.

In general, engineers design systems to achieve optimal performance $P_{max}$. However, this optimal value may not be attainable due to environmental complexity. The system typically operates with incomplete and uncertain information under limited computational resources. The term \textit{bounded rationality}, originally defined in decision theory \citep{Simon-1955}, describes this form of rational action under information and resource constraints. Consequently, in practice, the system tries to exceed a sufficient performance threshold, $P \geq P_{suf}$, which can be lower than the theoretical optimum, $P_{max} > P_{suf}$.

The measure $P$ allows for evaluating a system's degree of rational action from an external perspective, without needing to know its internal mechanisms. However, since the measure is specific to the task class $C$, this evaluation is not universal and, therefore, is not used as a general classifier, as argued in Section \ref{ref:approaches}. Its utility is limited to comparing systems that execute the same class of tasks. Thus, an external observer can compare the degree of intelligence of two systems (with respect to rational acting) by comparing the performance values each one achieves. In the case of two financial analysts, the first can be considered more intelligent than the second if the profit generated by their recommendations, $P_1$, exceeds that of the second, $P_2$ (i.e., if $P_1 > P_2$). Similarly, a chess player is generally considered more intelligent than another if the first wins a greater number of games.

\subsection{Learning}

To describe what is understood by learning, we can use the definition proposed by Tom Mitchell, based on system performance\footnote{This definition of learning as a behavior property that can be analyzed by an external evaluator contrasts with an internal view that understands learning as a process of knowledge acquisition, i.e., as the acquisition of new beliefs through processes of memorization or inference (e.g., induction or deduction).} \citep{Mitchell-1997}:

\begin{quote}
\textbf{Learning}. An agent learns from experience $E$ with respect to a class of tasks $C$ and a performance measure $P$ if its performance at tasks $T \in C$, as measured by $P$, improves with experience $E$.
\end{quote}

In general, experience $E$ includes results from executing multiple tasks $T \in C$. For example, chess players acquire experience through the favorable or unfavorable results of multiple games and use it to refine their ability to defeat their opponents, such that performance progressively improves with the number of games played. Depending on the type of information available in $E$, the system can follow different learning strategies. The system can use examples provided by an instructor with supervised learning (if the examples are labeled) or unsupervised learning (if they are not), or it can use positive or negative rewards with a reinforcement learning strategy.

To measure a system's learning capacity, an external evaluator employs the performance measure $P$. This can be used to measure the magnitude of the improvement $\Delta P = P' - P$, that is, the difference between performance after and before learning. It is also possible to measure the speed at which the system learns, i.e., the speed at which the system improves its performance until it reaches a sufficient performance threshold $P_{\text{suf}}$.

Different forms of learning can also be distinguished according to the way in which the system acquires experience $E$. It is called \textit{off-line} learning if the system has the complete set of examples at the beginning of the learning process, and \textit{on-line} learning if the system obtains the examples incrementally as it operates in its environment. The term \textit{continual learning} refers to the ability of a system to improve its performance throughout its entire operational life cycle, integrating the experience of each execution without degrading the performance achieved in previous stages \citep{Parisi-2019}. 

\subsection{Autonomy}

Autonomy is another property commonly associated with intelligent behavior \citep{Castelfranchi-1994, Wooldridge-1995, Franklin-1996, Wooldridge-2009, Shavit-2023}. It is a concept that allows for different definitions depending on the theoretical framework in which it is framed \citep{Boden-2008, Vernon-2014}. According to the operational context defined in this work, we can use the following definition:

\begin{quote}
\textbf{Autonomy}. An agent is autonomous with respect to a task $T$ if it is capable of translating by itself, without user intervention, the request to perform task $T$ into the execution of a sequence of actions $\langle a_1, a_2, \ldots, a_k \rangle$ that successfully fulfills such a request.
\end{quote}
Evaluating a system's autonomy admits partial degrees and different dimensions. For instance, according to the level of user intervention, autonomy could be graded according to the levels shown in Table \ref{tab:autonomy_levels}, which presents similarities to other scales proposed in the literature \citep{Sheridan-1978, Endsley-1999, Beer-2014}. This scale can be related to the common distinction between \textit{human-in-the-loop} systems, in which the user actively participates in the acting cycle (levels 1, 2, and 3), and \textit{human-on-the-loop} systems, in which the user maintains general supervision without intervening in each individual action (level 4), up to the fully autonomous performance of level 5.

\begin{table}[ht]
\centering
\footnotesize
\renewcommand{\arraystretch}{1.5}
\setlength{\tabcolsep}{5pt}
\begin{tabular}{l l p{6.5cm}}
\toprule
\textbf{Level} & \textbf{System Category} & \textbf{Description} \\
\midrule
\textit{Level} 1  & Information system & The system provides the user with information about the state of the environment and the user uses this information to decide and execute actions.\\ \addlinespace
\textit{Level} 2  & Recommender system & The system recommends how the user should act, and the user executes the actions in the environment.\\ \addlinespace
\textit{Level} 3  & Assistant system & The system acts in the environment, but each action requires confirmation from the user before being carried out.\\ \addlinespace
\textit{Level} 4  & Supervised autonomous system & The system acts in the environment under the supervision of the user, who may intervene if necessary.\\\addlinespace
\textit{Level} 5 & Fully autonomous system &
The user only specifies the task, and the system carries out the entire operation without any user intervention. \\
\bottomrule
\end{tabular}
\caption{Levels of autonomy according to the degree of user intervention.}
\label{tab:autonomy_levels}
\end{table}

The degree of a system's autonomy can be evaluated not only by the degree of human intervention it requires, but also by the diversity of  situations in the environment in which the system operates correctly (i.e., with performance $P \geq P_{suf}$). A system capable of executing tasks of a class,  $T \in C$, without human intervention under very diverse conditions exhibits a higher degree of autonomy than one that can only operate in highly controlled environments. For instance, a vehicle with a high degree of autonomy is able to complete a journey without driver intervention both on a clear highway and in a city center with dense and unpredictable traffic. To describe the degree of autonomy of driverless vehicles with respect to the diversity of situations in the environment, the \textit{operational design domain} (ODD) concept has been used \citep{Shakeri-2024}, which indicates the set of constraints (geographical, meteorological, traffic-related, among others) under which the system guarantees its correct operation.

It is important to note that an autonomous system may reject a task requested by the user because it conflicts with the system's own goals (i.e., goals permanently included by the developer), such as safety objectives or social norms (e.g., a passenger in an autonomous car cannot request that the vehicle ignore traffic laws). It may also reject a task due to limitations in the system's capabilities or the current state of the environment.

A desirable requirement related to autonomous behavior is the alignment of user and system goals so that both work toward the same objectives. In general, goals may present conflicts, as described by the AI value alignment problem \citep{Gabriel-2024} or the principal-agent problem \citep{Ross-1973}. Goal alignment is essential to ensure that the system operates in accordance with the user's ethical values, such as fairness (avoiding group bias) or limiting harmful use\footnote{An example of goal conflict related to the ethical behavior of autonomous vehicles has been analyzed by Bonnefon et al., considering extreme situations where vehicles must choose between hitting pedestrians to save their passenger or sacrificing the passenger to save pedestrians \citep{Bonnefon-2016}.}. This alignment issue connects to the broader field of AI governance, which, among other topics, examines the mechanisms needed (such as regulatory, organizational, and technical measures) to ensure that autonomous systems operate within limits acceptable to society, classifying systems according to their level of risk and establishing oversight requirements proportional to their degree of autonomy.

\subsection{Explainability}

The ability of intelligent systems to generate explanations about how they reach their conclusions has been emphasized in the context of what is known as explainable artificial intelligence (\textit{explainable artificial intelligence}) \citep{Doran-2017, Gunning-2019, Dwivedi-2023}. According to the operational context defined in this work, we can use the following definition:

\begin{quote}
\textbf{Explanation}. An explanation $X$ is a description generated by an agent that allows the user to understand what the agent relied on to produce the result $R$. 
\end{quote}

In this definition, a result $R$ refers to whatever the agent produces when executing task $T$. It may be an output that the user receives (for example, a classification, a diagnosis, or the recommendation of an action) or an action that the agent carries out (for example, moving a robotic arm, or braking an autonomous vehicle).

The explanation $X$  can consist of a feature importance attribution that identifies the input features that contributed most to the result $R$, as is the case with saliency maps in computer vision. For example, in a credit approval system, the explanation could show the degree to which debt and income level have been the most significant factors in the denial of a loan.

At a more elaborate level, the explanation $X$ can show a causal or logical chain that links the result $R$ to domain concepts, building a comprehensible step-by-step narrative. In the loan granting example, the explanation could describe that the loan was denied because the current debt level of 65\% exceeds the 40\% threshold established by the institution, which, combined with a monthly income below the 2,000-euro threshold, implies a high risk of default. The explanation can also take a counterfactual form \citep{Miller-2019}, answering questions like ``what if ...?'' to help the user understand what conditions would have led to a different result $R'$. For instance, if the debt level had been 35\% instead of the current 65\%, the loan would have been approved.

An agent's explanatory behavior can be evaluated externally by measuring how easily users understand the explanations provided by the system \citep{Hoffman-2018, Nauta-2023}. This understanding is estimated through objective metrics of syntactic complexity (for example, a more compact logical chain with fewer steps is usually more comprehensible than an extensive one) or through empirical studies with specific groups of users relevant to the tasks performed by the system (measuring, for instance, perceived trust in the explanations or the accuracy of user predictions regarding the agent's behavior).

An agent's explanatory behavior must also be evaluated according to its fidelity, i.e., the degree of precision with which the explanation describes the actual internal mechanisms used by the system to produce the results. However, this form of evaluation is not conducted from the external perspective of the user but from the perspective of the developer, who knows the system's internal processes. The importance of explanation fidelity has become particularly evident in language models, which can generate comprehensible explanations as chains of thought (\textit{chain-of-thought}) that, nevertheless, do not align with their internal logic \citep{Turpin-2023}.

Explanations can increase the user's trust in the system, which is especially important when the user is ultimately responsible for the actions to be executed. Explanations are also useful for identifying erroneous system behaviors (e.g., failures or unacceptable behaviors that do not align with human values). Due to its relevance, explainability has been included in regulations related to artificial intelligence \citep{Goodman-2017}, making this property a legal requirement in certain application contexts.

\subsection{Other properties}

This section summarizes other properties that have been related to the behavior of intelligent systems:
\begin{itemize}
\item \textit{Generality} refers here to the versatility of a system, i.e., performing multiple classes of tasks in different domains. For example, the Deep Blue system \citep{Campbell-2002} was designed exclusively to play chess, being incapable of performing other types of tasks. The concept of \textit{artificial general intelligence} (AGI) represents the goal of achieving a maximum degree of generality, comparable to human performance and versatility.
\item \textit{Proactivity} involves the system taking the initiative, operating in a goal-driven manner rather than merely responding reactively to user instructions or immediate changes in the environment \citep{Wooldridge-2009}. In the context of user interaction, proactivity is reflected in the anticipation of the user needs, giving rise to a form of interaction called mixed-initiative \citep{Allen-1999}, in which both the user and the system can take the initiative in communication. 
\item \textit{Adaptation} to changes in the environment is a property that has also been related to the behavior of intelligent systems \citep{Maes-1993,Brustoloni-1991}. At a first level, an agent acting rationally in a dynamic environment exhibits the ability to adapt, given that it continuously adjusts its actions to the environment's state with the purpose of maximizing a performance measure $P$. For example, an autonomous vehicle selects the route that minimizes travel time based on the traffic state at the start of the trip, and if it detects an unforeseen blockage during the trip, it modifies the route to continue minimizing the time. Adaptation also manifests at a second level if the agent learns continuously, as this allows it to prevent or mitigate the decrease in performance that changes in the environment can cause.
\item \textit{Argumentation} is another property that has been related to intelligent behavior in the field of AI \citep{Bench-Capon-2007}. An agent that uses argumentation generates a structured justification to defend the result $R$ of a task to the user using relevant and coherent premises, responding to objections to sustain that result. Argumentation is related to explainability, although both properties answer different questions. While explainability answers the question ``what did the agent rely on to generate the result $R$?'', argumentation answers the question ``why is the result $R$ correct or the most appropriate one?''.
\end{itemize}

Other properties that characterize the behavior of intelligent systems have also been analyzed in the AI literature, such as \textit{benevolence}, understood as the adoption of the goals of other agents without prior request \citep{Conte-1995}, the \textit{social ability} to interact with other agents \citep{Wooldridge-2009}, or \textit{veracity}, i.e., not deliberately communicating false information \citep{Evans-2021}.  Some of these properties, together with the autonomy and explainability discussed above, are receiving increasing attention in the field of AI governance, which seeks to establish principles, standards, and oversight mechanisms to ensure that intelligent systems operate within limits acceptable to society \citep{Shavit-2023}.

\section{Basic functions of an intelligent system} \label{ref:basic_functions}
This section describes the intelligent system's behavior (what the system does) from an internal functional perspective. According to the terminology used in systems engineering, such behavior is described as   \textit{internal functions} from which the primary function of the system emerges \citep{Crawley-2015}, or \textit{subfunctions} into which a global system function can be divided \citep{Pahl-2007}. Specifically, four basic functions are examined: perceiving the environment, evaluating the situation, deciding on actions, and executing actions.

\begin{figure}[htb!]
\begin{center}
\includegraphics[width=0.75\textwidth]{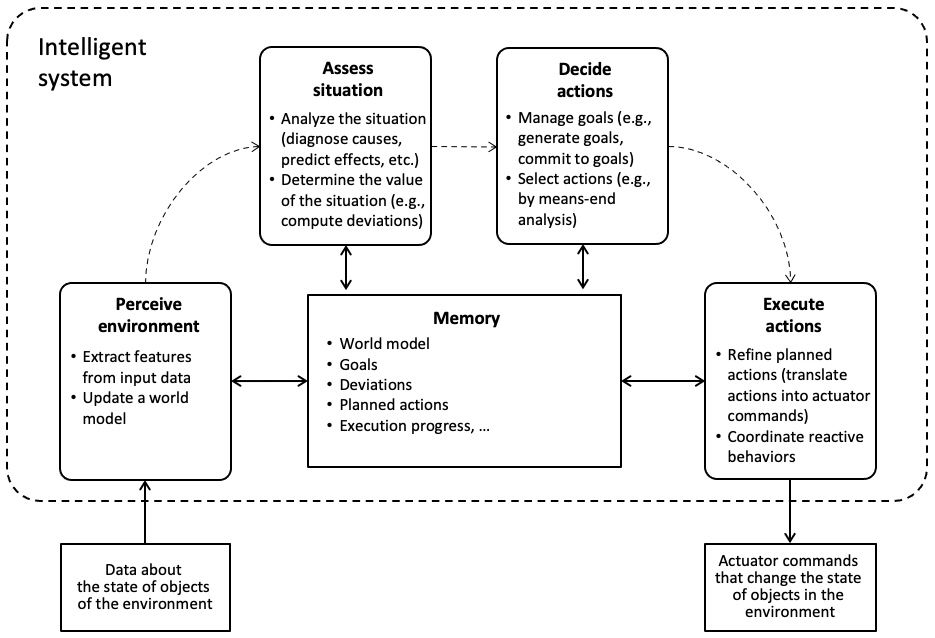}
\caption{Basic functions of an intelligent system (perceive environment, assess situation, decide actions, and execute actions) and their relation with input data on the state of the environment, commands to actuators, and the shared memory containing data used by the four functions. Solid arrows indicate data flow (input or output). Dashed arrows indicate the precedence order (idealized) in the execution of these functions.}
\label{fig:functional_decomposition_characterization}
\end{center}
\end{figure}

Figure \ref{fig:functional_decomposition_characterization} shows the way in which the four functions are organized in relation to the input data on the state of the environment, the output actuation commands, and a shared memory containing data used by the four functions. The figure also shows, by means of dashed arrows, the functional cycle \textit{perceive-assess-decide-execute}, which indicates the precedence order in executing these functions. This cycle should be considered an idealization of the usual precedence, given that certain systems in specific situations may not include some of these functions and the order in which they are executed may involve iterations involving, for example, separate decision-making and action-execution cycles (for replanning or error recovery).

This view summarizes similar functional decompositions described in the literature on software systems. For example, Albus identifies an architecture for real-time systems with four functional components (sensory processing, world modeling, value judgment, and behavior generation) separated into (1) situation assessment and (2) planning and execution \citep{Albus-1991}. The MAPE-K reference model for self-management software defines a similar architecture with four basic functions: monitor, analyze, plan, and execute \citep{IBM-2005}. In robotics, Ingrand and Ghallab identify similar functions in a context of deliberation: observation, goal reasoning, planning, acting, and monitoring\footnote{Ingrand and Ghallab, like other authors, also consider learning as a basic function of the system \citep{Ingrand-2017}. In this paper, learning is described as a property of behavior rather than a basic function, since learning (understood as improvement through experience) is a property also applicable to each basic function.} \citep{Ingrand-2017}. Murphy describes a canonical architecture for autonomous robots that includes deliberative functions such as generating plans, selecting resources, implementing plans (execution), and execution monitoring \citep{Murphy-2019}.

The cognitive analysis of human behavior has also identified similar functions. For example, Rasmussen describes a decision ladder \citep{Rasmussen-1974} with eight generic information processing steps in human decision-making, which is similar to the functional structure presented here. Lawson describes a command and control process with five basic functions: sense, process, compare, decide, and act \citep{Lawson-1981}. Endsley identifies the function of situational awareness (which includes perception, comprehension of the current situation, and projection into future states) in addition to the functions of decision and performance of actions \citep{Endsley-1990}. Some of these functions have also been described as capabilities of cognitive architectures. Langley identifies situation perception and situation assessment, prediction and monitoring, problem solving and planning, or execution and action \citep{Langley-2009}, and Vernon identifies perception and action selection, among others \citep{Vernon-2022}.

The following sections provide a more detailed explanation of each of these functions, including other sub-functions. In general, the description of each function is presented independently of its application domain and the logical mechanisms used for its implementation (understood as specific representation structures and algorithms). Nevertheless, in some cases, implementation examples are included for illustrative purposes.

\subsection{Perceiving the environment}

Perception is one of the essential functions of an intelligent system, as it allows the system to acquire information about the environment. An intelligent system can use sensors to automatically measure physical magnitudes about the state of objects in the environment (temperature, pressure, acceleration, light intensity, light wavelengths, etc.), including magnitudes related to the state of its own physical body (for example, battery charge, CPU temperature). The system can also receive information about the environment through user messages (or through messages from other agents) using a symbolic communication language\footnote{A distinction can be made between the interpretation of sensor data and the interpretation of messages that the user (or another agent) communicates to the system about environment data. In both cases, the system receives references to the characteristics of the environment. However, according to the terminology used by Pierce \citep{Deacon-1998}, sensor references can be understood as \textit{indices} and message references as \textit{symbols}. An index is a reference based on a causal relationship (for example, a temperature measurement is caused by a state of the environment), while a symbol is a subjective reference established by agreement between different agents (for example, the name of a city).}. For example, the user can provide a medical diagnostic system with information about the patient, such as age, sex, and the symptoms the patient describes, such as fatigue, difficulty breathing, etc.

Typically, perception updates a \textit{world model} \citep{Christensen-2016,Albus-1991,Burgard-2016}. This model is an approximation, potentially incomplete and uncertain, of the static and dynamic characteristics of the environment, including estimated properties that may not be directly observable. In the context of perception, the agent typically uses the world model to integrate and coherently maintain recent information about the variable state of objects in the environment that is relevant to the tasks performed by the system. 

According to this description, the function of the world model is to \textit{memorize} the state of the objects  of the environment perceived by the agent. However, the term world model is also used in the AI literature in a second sense, referring to a model whose function is to \textit{predict} how the state of objects in the environment would change if an action were executed \citep{Ha-2018}.

The following factors affect the complexity of perception: (1) heterogeneous perception, i.e., the system uses different types of sensors to observe the environment (e.g., acoustic sensors, lidar sensors, cameras, etc.), (2) uncertain perception, i.e., the system is not capable of observing all relevant properties of the environment to make decisions (due to, for example, insufficient sensors, measurement inaccuracy, the presence of noise in measurements, or the absence of sensors to measure some aspects of the environment), and (3) active perception, i.e., the system performs active data collection about the environment following a dynamic focus of attention.

The next sections describe in more detail two relevant functions that usually participate in the perception process.

\subsubsection{Feature extraction}

Data measured directly by sensors can contain a significant amount of information that is not necessary for the system's objectives. For instance, audio data collected by a microphone typically includes background noise that is not useful for speech recognition. Feature extraction involves identifying and isolating information from the raw data that is relevant to the tasks performed by the system. This process generates meaningful attributes or patterns called \textit{features} and helps reduce data dimensionality, eliminating noise without losing significant information. For example, features in images obtained by cameras could be edges, textures, or geometric shapes that are useful for object recognition.

Automated feature extraction allows an intelligent system to operate with a faster response time, as it can focus its attention on the aspects that influence decision-making. Furthermore, it provides greater robustness since the system's response is not affected by irrelevant changes occurring in dynamic environments (for example, variations in lighting in images or the presence of noise in audio recordings).

In certain problems, feature extraction requires considering the relationships between different elements of the input, since the interpretation of an observation may depend on its context (e.g., on other observations that occurred close in time or on nearby regions of an image). For example, in a computer vision system, a region of an image containing a round orange object can be interpreted as either an orange or a ball, depending on the elements appearing in other regions of the same image. Relating each observation to other elements in the input provides contextual information and helps refine its representation and interpretation\footnote{This function can be implemented using self-attention mechanisms, a central component of the Transformer architecture, which allows elements of an input to be related to one another and enables the construction of contextualized representations in the form of continuous vectors (\textit{embeddings}).}.

\subsubsection{World model update}

Intelligent systems that use a world model to internally represent the state of the environment and operate situated in a dynamic environment need to keep the model updated with respect to the changing state of the environment. This function consists of continually building and revising the model based on observed data so that it reflects the evolution of the environment over time \citep{Burgard-2016,Albus-1996}. The update is performed periodically at a sufficiently high frequency to allow the system to accomplish its tasks correctly.

In each update cycle, the system integrates newly acquired information with its existing information. The system must handle uncertainty when observations are incomplete or contain errors. In addition, the update process must preserve the coherence of the model, both at each instant and over time, ensuring, for example, that an object observed at different moments is recognized as the same entity.

At the implementation level, the world model is often described using representations adapted to the task and the type of environment. In complex environments, the model can combine multiple representations organized into hierarchies with various levels of abstraction. \citep{Pronobis-2012,Hughes-2022}. In such models, lower layers are closer to sensory data (e.g., images or point clouds), while higher layers represent more abstract information that is closer to the user’s language (such as objects, places, or categories). Updating this hierarchical model involves interactions between these levels, combining bottom-up processes of information aggregation with top-down processes of constraint\footnote{The process of updating the model using multiple modalities of environmental representation is related to \textit{perceptual anchoring} \citep{Coradeschi-2002}, which establishes a correspondence between symbols and attributes derived from sensory data through data structures known as anchors. In robotics, perceptual anchoring provides an operational solution to the problem of relating symbols to their meaning (\textit{symbol grounding}) \citep{Harnad-1990}.}.

\subsection{Situation assessment}

The assessment of the situation helps the intelligent system make informed decisions about which actions it should perform. Furthermore, it can help the user understand properties of the environment that are not directly observable. In this text, it is understood that situation assessment consists of two functions: (1) analyzing the situation in a broader context (e.g., projecting the present state into the future or analyzing possible causes) and (2) determining a value that expresses whether the situation is favorable or unfavorable. The following sections detail these two functions.

\subsubsection{Situation analysis}

The analysis of a given situation in a broader context typically includes the inference of unobserved beliefs, such as categories into which the situation is classified, states or events that are possible causes of the situation, or possible consequences of the situation. 

Situation analysis typically involves \textit{reasoning}, understood as the search of a sequence of valid inferential steps through which the agent, starting from facts it already considers true, can establish the truth of new facts about the state of the world. For example, an intelligent system in the medical domain can determine that the cause of a particular symptom is a specific viral infection by reasoning with the help of a set of general relationships between observable symptoms and infectious diseases. 

Situation analysis can also make use of world models with predictive functionality, i.e., models that anticipate how the state of objects in the environment may evolve. For example, an urban transportation management support system may predict the passenger demand for a bus line over the next few hours using a model that includes demand patterns according to the time of day and the type of date (weekdays, holidays, seasons of the year, etc.).

\subsubsection{Determining the value of the situation}

The situation assessment can be completed by determining values that indicate the degree to which the situation is favorable or unfavorable. The situation under assessment can be real or hypothetical and may include recent states and short-term predictions generated by the situation analysis described above. 

A way to determine the value of a situation is through functions that measure \textit{deviations}, quantifying the distance between the given situation and the desired situation. A desired situation may be expressed as a specific state (for example, a package must be delivered at a certain time, or a robot must have collected 50 units of a material) or as general conditions expressed as desired values for certain magnitudes (for example, computer temperature, battery charge, safe distance from obstacles, or acceptable vehicle acceleration to maintain passenger comfort).

The evaluation can include an additional assessment to generate \textit{drives} (or impulses) that quantify the importance that the agent assigns to such deviations. For example, a deviation from the desired battery charge can generate a drive value representing the urgency to recharge that a robot gives to the deviation. Using drives allows the same deviation to be weighted differently based on additional factors such as the relevance of the task in a mission, the character of the agent (audacious, cautious, etc.), or the agent's emotional state (frustration, satisfaction, anxiety, stress, etc.). The field of affective computing \citep{Picard-2000} provides a framework inspired by biological systems where drives and emotions are used to assess situations \citep{Arkin-1998} \citep{Canamero-2001} \citep{Marsella-2010} \citep{Moerland-2018}. 

\subsection{Deciding actions to take}

One of the main functions on which the autonomous behavior of an intelligent system is based is deciding what actions to execute in a specific situation. For example, an autonomous car decides the appropriate route to follow to reach a desired destination, considering traffic conditions, or a medical system decides which medication should be prescribed to a patient.

A factor that influences the complexity of the decision is whether the system's actions are \textit{sequential} or \textit{episodic}. Acting sequentially means that the effect of an action can affect future actions (for example, the movements of a robot in a room). Conversely, in the case of episodic actions, each action is independent of the previous ones, as occurs in an access control system that opens a door after recognizing a person's face. Sequential acting is more complex than episodic acting, since it requires anticipating the consequences of sequences of actions. Some authors have emphasized the capability to plan sequential actions as a relevant property of intelligent systems \citep{Chan-2023,Poole-2010}.

Another factor that affects the complexity of the decision is \textit{uncertain acting}. This occurs in dynamic environments where the system cannot predict the exact outcome of its actions. In these situations, the same action does not always guarantee the same effect, since the final state of the environment depends on external variables for which the system has no information. For example, the action of accelerating an autonomous car can lead to very different results due to unforeseen factors, such as a flat tire, an oil slick on the road, or a sudden mechanical failure. This lack of certainty forces the system to consider not only the most probable effect but also possible alternative effects.

In deciding which actions to take, we can distinguish the decision that determines \textit{what to do} (i.e., adopting goals) from the decision on \textit{how to do it} (selecting actions). The following two sections describe these two functions.

\subsubsection{Adopting goals}

An autonomous system can adopt goals that explicitly specify what the system must do. These goals can be classified into:
\begin{itemize}
    \item \textit{Conditions on states}. They express situations in the environment that must be met. For example, in an autonomous vehicle, a goal can indicate the destination. In this case, it is an \textit{achievement goal} that is discarded once achieved. There are also \textit{maintenance goals}, which define conditions that must be preserved over time and, therefore, are not abandoned after initial success \citep{Wooldridge-2009,Duff-2014}. For example, in a thermostat, the maintenance goal is defined by acceptable temperature intervals.
    \item \textit{Preferences to optimize}. They express how to quantitatively evaluate different courses of action. In the case of autonomous navigation, a goal of this type could indicate that the system should try to minimize the distance or time of the planned route. Unlike state conditions, preferences do not indicate when a goal has been satisfied, but rather serve to compare and select the best option among several alternatives.
\end{itemize}

The goals adopted by an agent can be external or internal. \textit{External goals} come from the tasks requested by the user of the system. For example, in an autonomous vehicle, the passenger can set goals such as the travel destination and the preference to minimize economic cost over travel time. On the other hand, \textit{internal goals} are generated by the intelligent system. These goals typically derive from criteria established by the developers (such as ensuring system safety or compliance with social norms), or they may arise as subgoals necessary to achieve a global goal. For example, if a warehouse robot detects a critical battery level, it will generate the subgoal of going to a charging station to ensure its own survival and to be able to continue with its main tasks.

When multiple goals exist, the agent must be able to resolve potential conflicts. The agent can give higher priority to goals using utility-based methods that calculate which goals offer the greatest benefit. Alternatively and more directly, the agent can use drive intensity (resulting from the evaluation of the situation) to select some goals over others. For example, a warehouse robot may decide to go to the charging station first instead of going to the delivery destination of a package, based on the drive that quantifies the urgency of charging the battery.

The \textit{belief-desire-intention} (BDI) model has been used in artificial intelligence as a reference for building intelligent agents that operate by generating and selecting goals \citep{Bratman-1988}. According to the BDI model, the agent can treat its chosen goals as \textit{intentions}, with which it persists over time instead of constantly recalculating at every moment. This allows the system to remain stable while executing complex plans, although it remains flexible enough to reprioritize if a high-urgency goal (such as a sudden obstacle or a critical system failure) takes precedence over the current mission.

In addition to generating and pursuing goals, it is necessary to monitor execution to determine when goals have been reached or when they should be abandoned. For example, a goal may be abandoned if it becomes impossible (e.g., a destination is blocked by a permanent obstacle) or irrelevant (e.g., a delivery robot is informed that the package is no longer necessary). The agent can also change goals through reprioritization when a new goal of higher priority arises.

\subsubsection{Selecting actions}

Action selection can be understood as a \textit{planning} problem whose goal is to find a sequence of actions to reach a desired state in the environment. Depending on the specific type of system and the type of actions to be taken, action selection can also be understood as another class of problem. For example, it can be understood as an \textit{assignment} problem that assigns needs to resources, \textit{scheduling} that organizes the temporal execution of a set of activities, or \textit{configuration} that finds how to assemble components to create a complex artifact that satisfies certain specifications. Table \ref{tab:reasoning_about_actions_examples} shows examples of these classes of problems for different types of systems.

\begin{table}[ht]
\centering
\footnotesize
\renewcommand{\arraystretch}{1.5}
\setlength{\tabcolsep}{5pt}

\begin{tabular}{p{2.7cm}p{3cm}p{7.5cm}}
\toprule
\textbf{Problem Class} & \textbf{System} & \textbf{Description} \\
\midrule
Planning &
Mobile robot &
A mobile robot considers possible routes to follow (using a map of the environment) and chooses the best route according to a performance measure (e.g., minimum travel time).\\ \addlinespace
Configuration &
Mechanical design aid system &
To design the machinery of an elevator, an intelligent system searches for a valid configuration of its components (cabin, cables, door, motor, etc.) taking into account the specifications provided by a customer.\\ \addlinespace
Assignment &
Airport terminal management system &
An intelligent system generates an assignment of boarding gates to flights by analyzing whether the possible assignments satisfy the airport's operational constraints.\\
\bottomrule
\end{tabular}
\caption{Examples of problem classes for selecting actions.}
\label{tab:reasoning_about_actions_examples}
\end{table}

Action selection is considered \textit{deliberative} when it meets the following requirements: (1) explicitly representing goals, (2) generating possible courses of action by predicting their effects on the environment, and (3) applying a criterion to select the best alternative. This process allows for changing goals, anticipating future situations, and acting in a justifiable manner. However, it consumes time and computational resources, which can make it unsuitable in dynamic environments that require rapid responses.

Alternatively, action selection is \textit{reactive} when the action is generated as an immediate response to a given situation, using, for example, situation-action patterns. Reactive selection is more efficient than deliberative selection, but it has the disadvantage that goals cannot be modified, as they are implicit, and the system lacks the ability to justify its decisions. For example, a medical system may decide a patient's treatment using symptom-treatment patterns learned from thousands of expert cases, but such patterns do not show the biological mechanisms of the drugs' effects on the body\footnote{In artificial intelligence, reactive behavior has been described in contrast to deliberation \citep{Wooldridge-2009,Murphy-2019}. The distinction between reactive and deliberative approaches has also been described in psychology by dual-process theories (e.g., \citep{Evans-1984,Kahneman-2011}). For example, Kahneman describes this duality using the names System 1 and System 2. System 1 (or intuitive thinking) is fast, automatic, unconscious, and non-linguistic (difficult to verbalize). In contrast, System 2 (or deliberative reasoning) is slow, logical, effortful, conscious, linguistic (can be described verbally), and makes complex decisions.}.

In the reactive approach, action selection is prefixed in a model for each class of situation (for example, through situation-action patterns). This forces the developer to anticipate all possible types of scenarios, which may not be feasible in complex environments\footnote{Machine learning techniques applied to action selection can facilitate the construction of reactive models that predict actions from situations (for example, policy models represented through neural networks). These models are trained using objective functions that then remain implicit in the learned model.}. Conversely, the deliberative approach prevents action selection from being prefixed. Instead, the system itself generates the best action in real time, evaluating candidate options to achieve its goals in each situation.

Note that the behavior of a reactive agent may appear rational when observed externally, but its internal operation does not follow a decision process based on maximizing a performance measure. In contrast, a deliberative agent that represents its goals as explicit preferences does operate internally according to a rational decision process, given that it selects the course of action that optimizes its performance measure. As previously indicated, this way of deciding actions is nonetheless limited by the system's computational capacity to generate combinations of action hypotheses and by uncertainty, when the system cannot anticipate with complete certainty the value of the performance measure for future actions.

\subsection{Executing actions}

The execution of actions typically involves translating action plans into executable commands and monitoring the environment to verify the success or failure of those actions. The following sections describe common functions that allow an intelligent system to execute actions.

\subsubsection{Refining actions}

The execution of actions may require a refinement process when the actions generated by the planning process cannot be executed directly. The action selection process often generates sets of actions in abstract form, leaving the specific details for the moment of execution. To execute these actions, it is necessary to translate them into detailed commands (e.g., commands interpretable by actuators of robotic systems). In certain cases, this refinement is carried out in continuous interaction with the environment to adapt the execution to the changing state of the environment.

An effective way to refine actions consists of employing a collection of components called \textit{skills} \citep{Bonasso-1997,Ingrand-2017,Mayr-2023} that encapsulate specific functional capabilities, such as landing, following a vehicle, or picking up an object\footnote{The term \textit{behavior} is also used by other authors to refer to a concept similar to skill \citep{Kortenkamp-2016,Sanchez-Lopez-2017,Murphy-2019}}. These components usually execute high-frequency closed-loop control, processing input data from sensors and generating output to actuators. For example, a skill for following a vehicle can operate by (1) reading input data from speed and distance sensors (which measure the distance to the vehicle to be followed) and (2) sending commands to driving devices (e.g., steering, braking, and acceleration) to maintain a preset distance. These components are typically implemented as control laws similar to those used in control theory\footnote{Note that skills could be understood as simple agents in a multi-agent system, following the idea of the recursive application of interaction patterns described in Section \ref{ref:operational_context}. In such a system, a high-level coordinating agent generates abstract actions and acts as the user of the agents that encapsulate skills. Each of these agents receives the abstract action as a task to be performed and executes it, perceiving its environment through sensors and generating direct commands to the actuators.} \citep{Michaud-2016}.

Since environments can be dynamic and unpredictable, the execution of each skill usually includes a constant monitoring process to ensure its function is performed as expected and to detect possible execution failures \citep{Gat-1999}. This supervision is carried out at different times during execution:
\begin{itemize}
\item \textit{Pre-execution verification}. Each \textit{skill} operates only under specific conditions of its environment. For example, landing can only be initiated if an aerial robot is flying, or a robot must be near a door before attempting to open it. Therefore, its preconditions must be verified before execution begins.
\item \textit{Supervision during progress}. During execution, the state of the environment is monitored to ensure the situation evolves correctly (e.g., verifying that the altitude decreases during a landing sequence or that the execution time does not exceed an expected time limit).
\item \textit{Outcome reporting}. Upon completion, supervision determines whether the action was a success or a failure. This information is generally used by other processes to continue with the execution of the next actions in the mission plan or to reconsider the execution of the plan when one of the actions has failed.
\end{itemize}

The execution of complex actions involves the coordination of multiple individual skills. A complex action (e.g., overtaking a vehicle) can be performed through a sequence of skills (approaching the front car, following the front car, changing lanes safely, passing the vehicle, etc.). Combined execution requires precise temporal sequencing, where each intermediate phase must reach specific success conditions before proceeding to the next. At the implementation level, this sequencing can be orchestrated in real time using representations especially appropriate for this purpose, such as finite state machines (FSM), behavior trees, or Petri nets.

\subsubsection{Coordinating reactive behaviors}

In the approach described in the previous section, skills are activated to execute actions selected by a central process (e.g., a central planner). However, intelligent systems can also execute actions in reflex mode, without centralized decision-making. In this mode, multiple \textit{reactive behaviors} (for example, behaviors to avoid obstacles, land in case of emergencies, etc.) are executed concurrently, and their combined operation is controlled by certain coordination mechanisms \citep{Murphy-2019}. These reactive behaviors carry out high-frequency closed-loop control (like the skills described above), processing input data from sensors and generating outputs to actuators. 

Coordination mechanisms include methods such as \textit{subsumption}, where the outputs of high-priority behaviors (such as emergency landing) override the outputs of lower-priority behaviors. The intelligent behavior of a system can emerge from the interaction of simple reactive behaviors without the need for a central planner\footnote{The development of systems with architectures that use only reactive behaviors is the basis of Nouvelle Artificial Intelligence, a movement started in the mid-to-late 1980s \citep{Brooks-1986,Brooks-1991}.}.

The coordination of reactive behaviors can also use cooperative methods that, for example, blend different outputs into a single action, adding them proportionally. The intensity (or gain) of each specific reactive behavior can be dynamically modulated by the strength of drives derived from situation evaluation, which fluctuate according to sensory input \citep{Canamero-2001}. For example, in a mobile robot, a repulsion drive can move the robot away from obstacles. This drive can be blended with an attraction drive that pulls the robot toward a target (e.g., a light source), and the combination of both drives, according to their intensity, can generate the compound movement.

In general, modern complex intelligent systems use hybrid solutions that integrate the two ways of acting described above. Certain forms of acting are executed through concurrent reactive behaviors, while others are directed by a centralized decision-making process (e.g., using a central planner). In the development of intelligent systems, combining both approaches is a non-trivial problem that requires specific design decisions. In certain situations, for example, a deliberative planner should inhibit reactive decisions when immediate responses must be postponed to achieve better long-term goals.

\section{Architectural form of an intelligent system} \label{ref:architectural_form}

\begin{figure}[htb!]
\begin{center}
\includegraphics[width=0.65\textwidth]{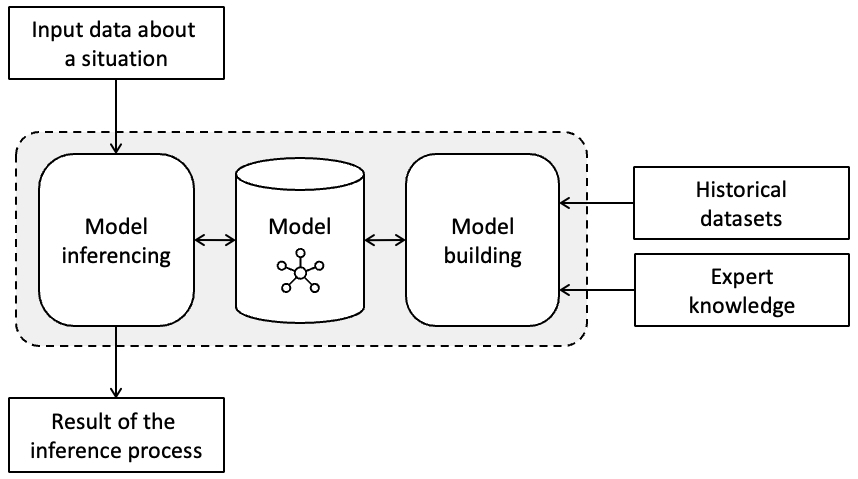}
\caption{The software architecture of an intelligent system typically includes components with a model-based structure, integrating a model (e.g., a neural network or a set of rules) along with processes for its inference and construction.}
\label{fig:ai_model}
\end{center}
\end{figure}

The software architecture of an intelligent system typically includes components with a model-based structure (see Figure \ref{fig:ai_model}) consisting of the following elements\footnote{Description adapted from \citep{OCDE-2022}.}:
\begin{itemize}
\item The \textit{model} captures relevant information from a given domain through a computational representation. This representation can take various forms, such as a neural network used for medical image diagnosis, a set of heuristic rules used to guide investment decisions, or an urban map for route planning.
\item The \textit{model inferencing} process refers to the automatic procedures that perform tasks in each given situation (diagnosing a disease, planning a route, etc.). These procedures usually implement general inference algorithms, such as automated deduction in predicate logic, feedforward propagation in neural networks, or belief propagation in Bayesian networks.
\item The \textit{model building} process refers to the procedures used to generate the model, which can be automatic (e.g., through machine learning algorithms) or manual (e.g., through the elicitation and analysis of human expert knowledge).
\end{itemize}

This modular structure, which separates the model from the construction and inference methods, allows for more effective handling of complex system development. For example, developers can use software tools with general machine learning and inference algorithms, independent of each particular domain, to build and use models in different applications. Machine learning tools also allow the model to be updated to respond to new requirements without the need to modify the inference procedures.

\subsection{Representation and inference}

The representation used by models follows two main approaches in artificial intelligence:
\begin{itemize}
\item \textit{Symbolic} approach: models represent what an external observer understands that the system knows about a domain of application, using formalisms such as knowledge graphs or rule-based systems\footnote{This approach is based on what Allen Newell and Herbert Simon identified as the \textit{physical symbol system hypothesis}, which maintains that a physical symbol system has the necessary and sufficient means for general intelligent action \citep{Newell-1976}.}.
\item \textit{Neuro-computational} approach (or \textit{connectionism}): models use representations in the form of networks inspired by the neural structures of biological systems.
\end{itemize}

Both approaches differ in the interpretability of their representations. Symbolic models employ symbols that represent elements of the environment (e.g., objects, events, or attributes) whose meanings are directly understandable to the user, as the user employs the same symbols to describe them. In contrast, the neuro-computational approach uses networks with quantitative parameters (numerical weights) obtained through machine learning. These values do not correspond directly to the symbols used by humans, so their meanings are not directly interpretable.

Within symbolic representation methods, two main paradigms are usually distinguished. The \textit{logic-based} paradigm employs representations with formal semantics from mathematical logic (e.g., rule-based systems, logic programs, description logics, and constraint-based representations). The \textit{probabilistic} paradigm addresses the representation of uncertainty and inference with solutions based on probability theory. Within this category are probabilistic graphical models, such as Bayesian networks. There are also hybrid or extended representations, such as Markov logic and fuzzy logic.

The suitability of these methods depends, among other factors, on the classes of functions they must perform. Table \ref{tab:examples_ai_methods} shows examples of these methods according to the functional decomposition presented in Section \ref{ref:basic_functions}. Neuro-computational methods are typically used to implement functions related to sensorimotor abilities that humans perform unconsciously, whereas symbolic methods have traditionally been applied to consciously accessible cognitive capabilities, such as diagnosis or planning. 

However, this correspondence has become less clear in recent years. The widespread use of large language models (LLMs) based on neural architectures has shown capabilities potentially applicable to functions traditionally associated with the symbolic approach. However, language models have limitations in terms of reliability and response time, as well as high computational resource consumption. For this reason, symbolic approaches may be preferable in critical scenarios where reliability, response speed, and resource efficiency are essential requirements.

\begin{table}[ht]
\centering
\footnotesize
\renewcommand{\arraystretch}{1.5}
\setlength{\tabcolsep}{5pt}
\begin{tabular}{p{3cm}p{5.5cm}p{6.5cm}}
\toprule
\textbf{General functions} & \textbf{Symbolic methods} & \textbf{Neuro-computational methods} \\ 
\midrule
Perceiving the environment &
Scene graphs, labeling via constraint satisfaction. &
Convolutional neural networks, attention-based models (transformers), multimodal models.\\  \addlinespace
Assessing the situation &
Rule-based systems, predicate logic, Bayesian networks, ontologies. &
Language models with reasoning capabilities for classification or diagnosis.\\  \addlinespace
Deciding actions &
Hierarchical task network (HTN) planning, constraint-satisfaction-based planning, temporal action logics, state-space search.&
Language models with reasoning capabilities for planning, internal simulation with world models for planning.\\  \addlinespace
Executing actions &
Finite state machines, behavior trees, Petri nets, fuzzy controllers. &
Neural policy networks, generative action models.\\ 
\bottomrule
\end{tabular}
\caption{Examples of symbolic and neuro-computational representation and inference methods used in building intelligent systems.}
\label{tab:examples_ai_methods}
\end{table}

\subsection{Model construction}

The process of building the models of an intelligent system can be carried out in two ways:
\begin{itemize}
\item \textit{Knowledge-driven} construction: the model is built from the knowledge of human experts. For example, in medical diagnosis, experts provide criteria relating symptoms to diseases, which can be modeled through symbolic representations (such as rules). This process is mainly manual, although semi-automatic solutions exist to help structure such knowledge.
\item \textit{Data-driven} construction: the model is generated automatically by applying machine learning algorithms to datasets. For example, a neural network can be trained with thousands of radiological images labeled as positive or negative examples.
\end{itemize}

Knowledge-driven construction was the dominant paradigm until the 1990s. However, since the manual acquisition of expert knowledge requires significant effort, its use is restricted to the development of models of limited size (this limitation is known as the knowledge acquisition bottleneck)\footnote{Ontology engineering proposes common representations that facilitate the exchange and reuse of knowledge bases, mitigating this problem.}. In contrast, data-driven construction has gained popularity thanks to increased computing power and the availability of massive datasets in various formats (images, unstructured text, time series, etc.).

It is worth noting that data-driven model construction is applicable to both neuro-computational and symbolic representations. For example, similarly to how a neural network can be trained to identify pathologies from multiple examples (e.g., tumor images), the model of a rule-based medical expert system can be generated through inductive learning algorithms that extract knowledge from historical medical records.

Machine learning can be applied in two different phases of an intelligent system's lifecycle. On the one hand, the automatic construction of a model can take place during the system's \textit{construction phase}. This allows developers to validate the quality of the model in a controlled environment before deploying the system for use. On the other hand, the model can also be updated through automatic methods during the \textit{operation phase}, i.e., while the already-deployed system is in use. However, this second option carries the risk of the system learning in an inadequately controlled manner, for example, from biased or adversarially manipulated data\footnote{This problem occurred with Tay, Microsoft's chatbot from 2016, which developed racist biases after interacting with uncontrolled public data.}, or by adapting improperly to changes in the underlying data distribution. For this reason, when this strategy is adopted, it is common practice to incorporate human oversight mechanisms (\textit{human-in-the-loop}) that allow model updates to be monitored and validated before they affect the system's behavior.

In this context, the validation of models with neuro-computational representations presents difficulties derived from their lack of interpretability. Neural networks behave as black boxes, making it difficult to understand how the model arrives at a given conclusion. For this reason, their validation must rely on specific methods such as those developed in the field of explainable AI (XAI) \citep{Phillips-2020}. Among other things, these techniques make it possible to identify which input features have the greatest influence on the model's predictions and to detect potential unwanted biases.

\subsection{Component integration}

The complete architecture of an intelligent system usually combines multiple model-based components with different representations and inference methods. Complex intelligent systems can also include components that do not have a model-based structure or whose models use representations outside the paradigms of artificial intelligence (for example, parameters of statistical regression functions or differential equations describing the dynamics of physical systems).

The way of organizing and integrating these components has evolved significantly from the earliest knowledge-based systems to today's architectures. In the 1990s, the integration of model-based components was addressed mainly from the field of knowledge engineering. One line of work proposed explicitly separating the domain model from reusable problem-solving methods (PSMs), specialized in tasks such as diagnosis, planning, or resource assignment, with the aim of facilitating their reuse across different applications. Representative examples of this approach are the CommonKADS methodology \citep{Schreiber-2000} and the development environments Protégé-II \citep{Eriksson-1995} and MIKE \citep{Angele-1998}. In parallel, the KSM environment \citep{Cuena-1997} proposed an organization in which the model-based component constituted the basic modular unit, with its integration described through specific languages.

In robotics, component integration is usually based on specialized modules that implement functions such as perception, localization, navigation, planning, and control. Middleware platforms such as ROS \citep{Quigley-2009} and its evolution, ROS 2 \citep{Macenski-2022}, provide the infrastructure needed to organize these modules as independent components that exchange information through messages. Reusable components for specific functions have been developed on top of these platforms, such as Nav2\footnote{\url{https://docs.nav2.org}} for autonomous navigation or MoveIt\footnote{\url{https://moveit.ai}} for motion planning and execution, as well as modular architectures based on the concept of skills or robotic behaviors, which facilitate the composition and reuse of robotic capabilities \citep{Mayr-2023,Fernandez-Cortizas-2023}.

Complex neural architectures also constitute a form of integration of multiple specialized components, implemented as subnetworks or neural blocks that perform differentiated functions, such as those described in Section \ref{ref:basic_functions}. A representative example is a vision-language-action (VLA) model, which integrates modules for perception, planning, and action generation \citep{Black-2024}. This type of architecture can be trained end-to-end, allowing internal representations between functional modules to be learned automatically during training. This approach offers great flexibility for optimizing the overall behavior of the system, although it usually requires greater computational resources and makes the interpretation of the learned internal representations more difficult.

More recently, the generalization of language models has given rise to a new approach for integrating specialized components. In these architectures, functions such as perception and action execution are usually delegated to specific tools, while a language model carries out situation assessment and action planning. This organization provides flexibility for incorporating and combining heterogeneous tools and knowledge sources, as the language model can dynamically decide which components to use and how to sequence their use depending on the task and the information obtained during execution. These architectures can be implemented through orchestration platforms that allow the workflows between the different components to be defined and integrated.

Taken together, the above examples show that the integration techniques have evolved significantly, although the modular organization of model-based components according to specialized functions (such as perceiving the environment, assessing the situation, deciding on actions, and executing actions) remains one of the most widespread architectural strategies for designing complex intelligent systems. 

\section{Examples of intelligent systems} \label{ref:examples}
This section shows examples of intelligent systems that are analyzed using the conceptual framework described in this paper. Various categories of systems are shown, and within each of them, a particular example of a documented real system is detailed. For each example, the following information is provided:
\begin{itemize}
\item \textit{Operational context.} The operational context of the system is described, indicating who the user is, what the work environment is like, and the characteristics of the interaction.
\item \textit{Intelligent behavior}. The properties that characterize the intelligent behavior of the system are described by indicating: the performance measure that quantifies the degree to which the system acts rationally, the degree of autonomy of the system, the capacity of the system to improve its performance with experience through learning, and the capacity of the system to explain to the user how it reaches its conclusions.
\item \textit{Basic functions.} This information provides details about the system's capacity to perceive the environment, assess the situation, decide on actions, and execute actions. 
\item \textit{Architectural form}. The model-based components that are part of the architecture of each system are described. This includes the representations used in the models, along with their inference and construction processes.
\end{itemize}

\subsection{Autonomous robots}

An autonomous robot is a representative case of an intelligent system that usually includes most of the properties described in this paper. In this category, there are systems with different degrees of complexity, such as industrial collaborative robots, autonomous vehicles (for example, self-driving cars or vehicles for planetary exploration), assistive robots, domestic cleaning robots, entertainment robots, etc.

\begin{figure}[htb!]
\begin{center}
\includegraphics[width=0.55\textwidth]{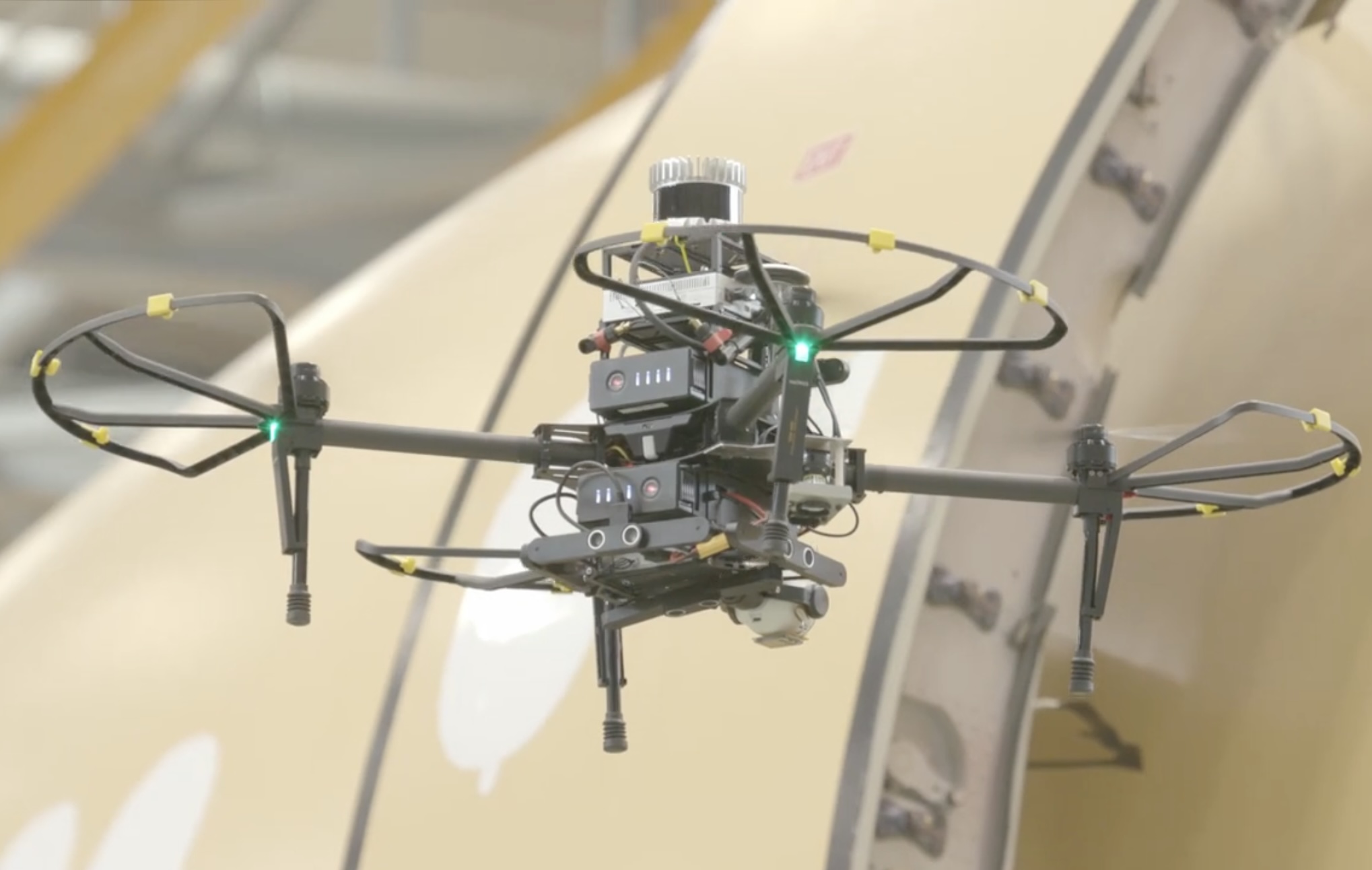}
\caption{Aerial robot for aircraft inspection (source: \url{https://vimeo.com/393907228}).}
\label{fig:inspection_foto}
\end{center}
\end{figure}

An example of this type of system is an aerial robot for aircraft inspection \citep{Bavle-2019}. The objective of this robot is to help maintenance engineers inspect the surface of an aircraft to locate possible defects. The robot is capable of flying close to the surface of the aircraft and processing images obtained with the help of on-board cameras.

The body of the robot is a multi-rotor aerial platform (Figure \ref{fig:inspection_foto}) equipped with multiple sensors, such as (1) an inertial measurement unit (IMU) to provide information on orientation, angular velocities, and linear accelerations, (2) cameras to calculate speed through visual odometry, (3) a LiDAR sensor that generates a 3D point cloud, and (4) a high-resolution front camera.

\begin{figure}[htb!]
\begin{center}
\includegraphics[width=0.70\textwidth]{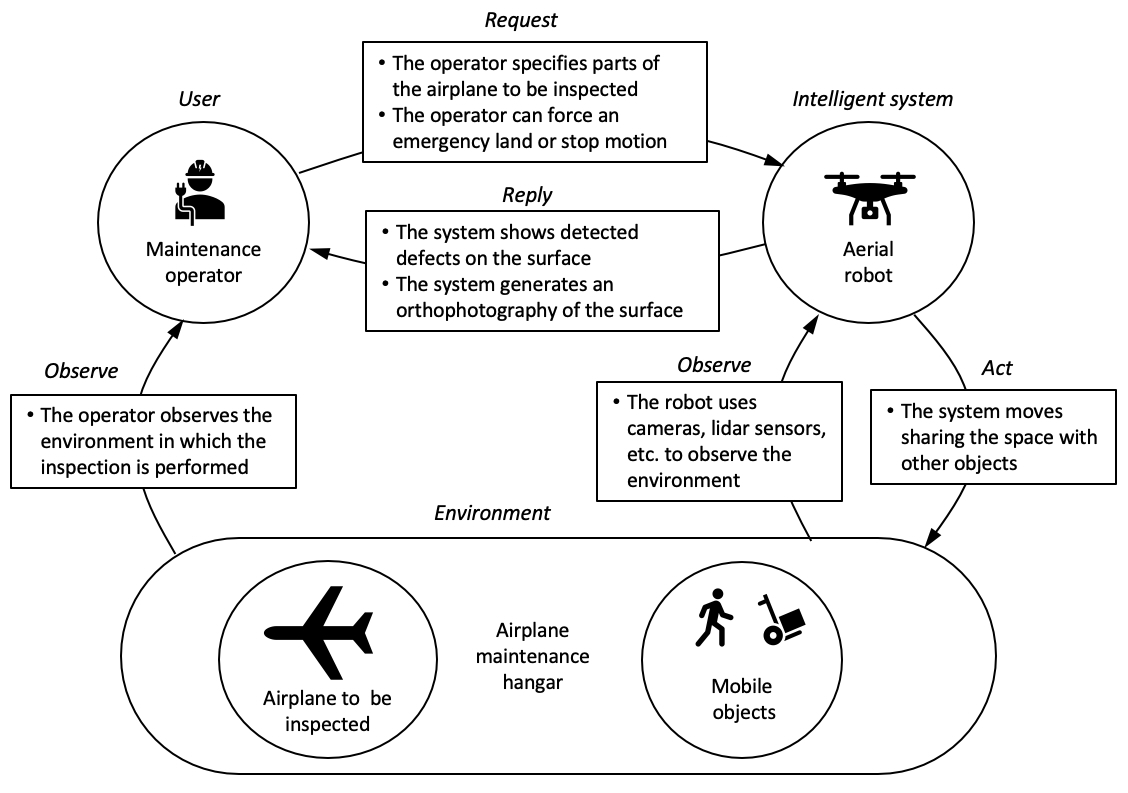}
\caption{Operational context of the aerial robot for aircraft inspection.}
\label{fig:inspection_example}
\end{center}
\end{figure}

Figure \ref{fig:inspection_example} shows one way to understand the operational context of this system. The user is the maintenance operator who uses the aerial robot to inspect the aircraft. The environment is the maintenance hangar with the aircraft to be inspected.

The properties related to the intelligent behavior of this system are as follows:
\begin{itemize}
\item \textit{Rational acting.} Rational action can be evaluated using a combination of the following performance measures: maximizing the area of the inspected surface, maximizing the accuracy of image classification, and minimizing inspection time.
\item \textit{Autonomy:} The system operates in a real-world environment without restrictions and, according to the levels of autonomy described in section \ref{ref:behavioral_properties}, the system exhibits level 4 (supervised autonomous system) because it performs the inspection task by itself, although under user supervision (the operator can intervene if an unexpected event occurs).
\item \textit{Learning.} This system does not learn \textit{online} during the operational phase, while the system is in use, and therefore cannot improve its performance through repeated inspection tasks. The system learned \textit{offline} during its construction phase to recognize defects using computer vision.
\item \textit{Explainability}. The operator can view graphs and images on the ground station screen showing the environment as perceived by the robot (point clouds and surface defects) and the planned route. In addition to this information, the robot does not provide any explanation of how it reaches its conclusions.
\end{itemize}

The basic functions performed by this system are the following:
\begin{itemize}
\item \textit{Perceive the environment}. The robot observes its environment with the help of the sensors described above. The aerial robot uses a three-dimensional map of the environment and localizes its position and orientation on the map by fusing data from multiple sensors using an extended Kalman filter (EKF). The robot generates an orthophotograph of the aircraft surface.
\item \textit{Assess the situation}. The robot recognizes anomalies on the surface of the aircraft through computer vision and localizes them spatially. The system also assesses the risk of collision with obstacles and the battery charge.
\item \textit{Decide actions}. The robot generates a three-dimensional trajectory that ensures complete coverage of the aircraft's surface during inspection. 
\item \textit{Execute actions}. The robot uses motion control mechanisms to execute the navigation plan. The robot also includes reactive behaviors that stop its flight to avoid colliding with moving obstacles or to land immediately when the battery charge is too low.
\end{itemize}

The software architecture of the robot includes the following main model-based components:
\begin{itemize}
\item \textit{Neural network for computer vision}. This is a neural network specialized in computer vision for anomaly detection, using images of the aircraft surface as input. This network is trained during the construction phase through supervised machine learning with the help of a labeled dataset.
\item \textit{Aircraft model for path planning}. The robot represents relevant locations for inspection on a three-dimensional map, along with a point cloud representing detected surfaces. The robot uses this model to generate a route through a general path planning algorithm. The user provides part of the aircraft model (main locations to be inspected) and the rest is generated from sensor data.
\end{itemize}

\subsection{Decision support for the management of dynamic systems}

The management of complex dynamic systems (in domains such as energy consumption, public transport, environmental pollution, etc.) usually employs sensor networks that capture and transmit large volumes of data in real time\footnote{This type of scenario has been described in technological ecosystems such as \textit{smart cities} and \textit{smart territories}, and in more specific sectors such as \textit{smart grids} for energy management, the digitalization of industrial processes (Industry 4.0) and intelligent transportation systems (ITS).}. In this context, intelligent systems can process such data efficiently to help managers detect and diagnose incidents, as well as to decide the best action strategies. This type of system has been applied, for example, to airport incident response \citep{Jo-1997}, early response to flood emergencies \citep{Molina-2003}, and the supervision of electrical energy networks \citep{DaSilva-2012}.

\begin{figure}[htb!]
\begin{center}
\includegraphics[width=0.70\textwidth]{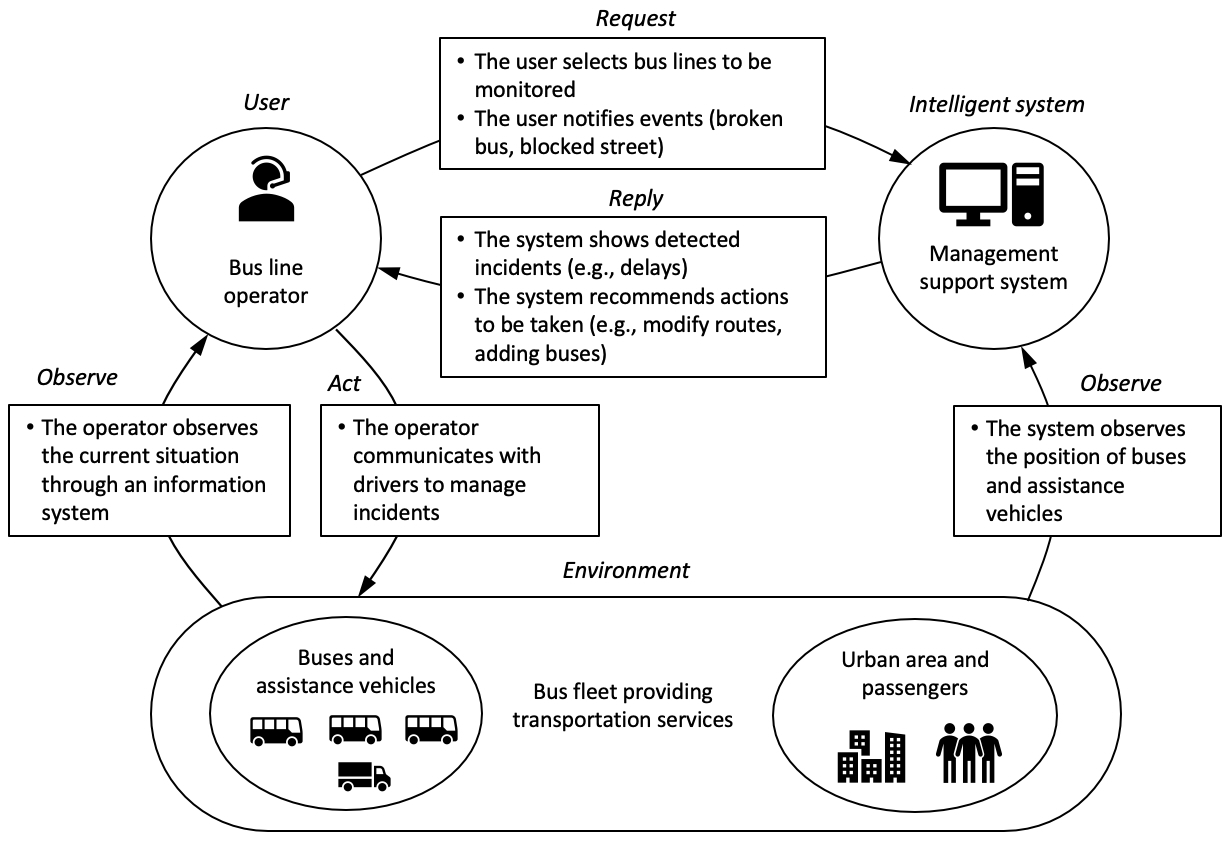}
\caption{Operational context of the intelligent system for support in public transport management.}
\label{fig:bus_transport_example}
\end{center}
\end{figure}

This section analyzes an example of this type of system in the public transport domain \citep{Molina-2005}. It is an expert system that was developed to detect incidents in an urban bus transport network and recommend actions to operators to resolve or mitigate the incidents. Figure \ref{fig:bus_transport_example} shows a way to understand the operational context of this system. In this case, the user is the bus line operator, and the environment corresponds to the fleet of buses that provides the transport services. The system periodically observes the position of the buses with an appropriate frequency to be able to respond quickly to the presence of incidents.

The properties related to the intelligent behavior of this system are the following:
\begin{itemize}
\item \textit{Rational acting}. Rational action can be evaluated using the following performance measure: minimizing travel time delays in the presence of incidents.
\item \textit{Autonomy}. According to the levels of autonomy described in section \ref{ref:behavioral_properties}, the system is classified as level 2 (recommendation system) because it recommends actions, but it is the operator who communicates these actions to the vehicle drivers.
\item \textit{Learning}. This system does not learn, since its performance does not improve with the repetition of the decision-support task.
\item \textit{Explainability}. The system justifies the recommended actions, showing the general strategy used to respond to the detected incident and how it prioritizes the actuation on certain bus lines over others based on predicted demand.
\end{itemize}

The basic functions performed by this system are the following:
\begin{itemize}
\item \textit{Perceive the environment}. The system receives real-time location data from the buses via GPS sensors and a communications network. The system uses a map as a world model that represents the characteristics of the transport lines and updates it periodically with the positions of the vehicles. The operator can communicate certain events to the system that the system cannot detect with the sensors (for example, a broken-down bus, a blocked street, etc.).
\item \textit{Evaluate the situation}. The system detects the presence of deviations from the transport plan and evaluates the severity of the delays detected on each line using the predicted bus demand. The system predicts short-term bus demand, taking into account the current time and date.
\item \textit{Decide actions}. The system determines the actions that must be carried out to manage incidents, proposing, for example, adding buses to lines with delays, changing blocked routes, or sending maintenance vehicles to repair broken-down buses.

\end{itemize}

The software architecture of the system includes the following model-based components:
\begin{itemize}
\item \textit{Transport lines}. The system uses a model in the form of a map with the characteristics of the transport lines (including stops, distances, transfer points, etc.) and the positions of buses. This model includes parking areas for maintenance vehicles and reserve buses.
\item \textit{Bus demand}. This model includes statistical measures on bus demand, such as the average demand on different days and hours. The function of this model is to predict bus demand for the next few hours. The content of this model was generated automatically during the construction phase of the system using historical data on bus demand.
\item \textit{Strategies for managing incidents}. This model consists of a set of heuristic rules that describe strategies for managing incidents (for example, adding more buses to a line with delays, sending maintenance vehicles, etc.). The recommended actions are generated using an automatic planner based on hierarchical task network (HTN) planning . The content of the model was formulated manually using information provided by domain experts.
\end{itemize}

In general, a decision support system, such as the one described above, uses a sensor network to perceive the behavior of the environment and includes mechanisms to evaluate the situation and recommend the actions to be taken. In this scenario, the human manager is responsible for making the final decisions and executing the actions in the environment. Therefore, it is essential that these systems communicate their recommendations with appropriate justifications to help managers assume responsibility for the decisions.

\subsection{LLM-based agents}

Another type of application that can be analyzed under the conceptual framework presented in this paper is the LLM-based agent, which has gained popularity in recent years \citep{Xi-2025}. This type of system employs a large language model (LLM) as a core component. 

The use of the language model allows users to interact with the system in natural language. The user can use natural language to request the system to perform tasks, and the system can also present its results in the same form. For example, the user can ask ``should I take an umbrella for my trip to Madrid tomorrow?'' and the system can answer ``no umbrella needed; the forecast for Madrid shows clear skies''. The user can also use other communication modalities (e.g., camera images) when the language models support them. 

The agent uses tools to extend the capabilities of the language model beyond text generation. These tools often include computational resources to access information (e.g., through web browsers or database access), execute code (e.g., using language interpreters), or use specific local or remote applications (e.g., calendars, email services, project management tools) by connecting through specific interfaces or standard protocols, such as the MCP (model context protocol\footnote{\url{https://modelcontextprotocol.io}}).

The agent also uses the language model (1) to automatically generate action plans that describe how to use available tools to address the requested tasks and (2) to analyze the outputs produced by these tools.  For example, continuing with the previous example, the agent could generate the phrase: ``I'm going to consult a meteorological service via the Internet to view the forecast for tomorrow''. This process can be interpreted, by analogy, as an inner monologue in natural language \citep{Huang-2022}, similar to the way people mentally verbalize strategies for carrying out a task\footnote{This form of operation is described, for example, by the ReAct model (Reasoning + Acting), which employs the LLM to generate integrated reasoning traces and specific actions \citep{Yao-2023}.}. 

LLM-based agents have demonstrated their utility in applications such as automatic code generation, customer service chatbots, and research assistants. However, these agents inherit the limitations of language models, such as unreliability (e.g., due to hallucinations), slow response times, and high computational costs.

\begin{figure}[htb!]
\begin{center}
\includegraphics[width=0.70\textwidth]{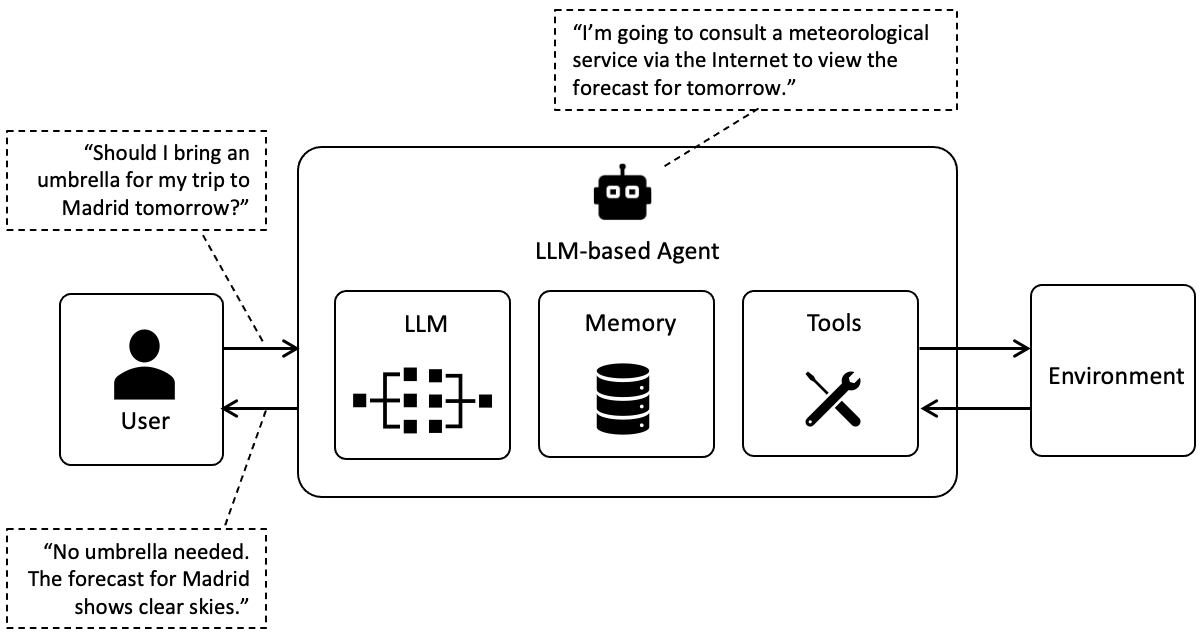}
\caption{Components of an LLM-based agent.}
\label{fig:llm_agent_components}
\end{center}
\end{figure}

The standard architecture of these agents includes the language model, a memory, and a set of tools to interact with the environment (Figure \ref{fig:llm_agent_components}). Memory helps the agent operate consistently over time by providing information obtained previously (e.g., previous conversations with the user and recent results of interactions with the environment). 

A specific category of this class of system is the agent for computer use (ACU) \citep{Sager-2025}. This agent is designed to execute tasks on a digital device (e.g., a desktop computer or a mobile phone) in response to instructions given in natural language. Within this category of systems is the vision-based agent for computer use, which uses a multimodal language model (with computer vision capabilities) to observe the computer screen and generate actions analogous to those performed via mouse or keyboard. Various research studies in AI proposed preliminary systems of this type, such as WebVoyager \citep{He-2024}, MobileAgent \citep{Wang-2024}, MobileAgent-v2 \citep{Wang-2024b}, and ScreenAgent \citep{Niu-2024}.

\begin{figure}[htb!]
\begin{center}
\includegraphics[width=0.70\textwidth]{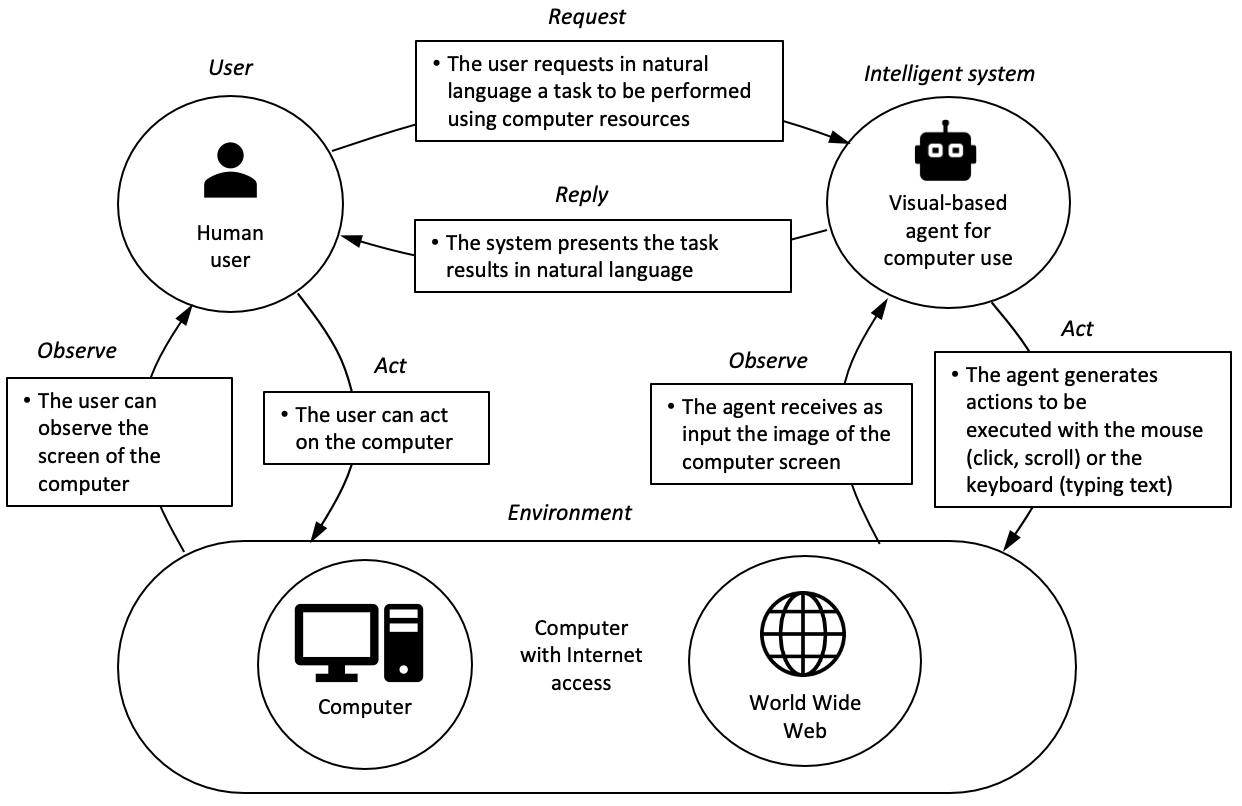}
\caption{Operational context of the vision-based agent for computer use.}
\label{fig:llm_agent_example}
\end{center}
\end{figure}

Figure \ref{fig:llm_agent_example} shows a way to understand the operational context corresponding to a vision-based agent (similar to WebVoyager) interacting with a web browser. The user communicates with the system in natural language to request a task that can be executed with the computer resources, and the system responds to the user in natural language with the result of the task. The environment is the computer used by the agent, including the Internet connection to access web-based applications. The agent observes the computer screen as an image (the agent may also receive auxiliary textual information extracted from the website's HTML code). The agent generates actions corresponding to mouse and keyboard operations, such as scrolling, clicking links or buttons, and entering text into an input box.

The properties related to the intelligent behavior of this system are as follows:
\begin{itemize}
\item \textit{Rational Action}. Rational action can be evaluated by measuring how successfully the system completes tasks requested by the user (e.g., using evaluation resources such as the WebVoyager benchmark \citep{He-2024}). The performance measure can be expressed as (1) maximizing the success rate of tasks and (2) minimizing the time or number of actions required to complete tasks.
\item \textit{Autonomy}. The system is conceived to operate on a computer in a fully autonomous way (level 5 according to the levels described in section \ref{ref:behavioral_properties}). However, the WebVoyager system is a demonstrative system with technological limitations that require further improvements in order to operate fully autonomously.
\item \textit{Learning}.  During the construction phase, the language model is automatically generated through pre-training and fine-tuning processes typically used for this type of model. During the operation phase, the language model’s parameters do not change while tasks are being executed. However, storing recent experiences in memory is a form of on-line learning, as it can improve the system’s performance when executing subsequent tasks.
\item \textit{Explainability}. The system can present natural language sentences corresponding to the chain of thought followed by the system, which can help to understand how it decides on actions and interprets their results. However, these explanations should not necessarily be interpreted as a faithful representation of the internal mechanisms involved, which are based on the use of neural networks and whose parameters are not directly interpretable by the user.
\end{itemize}

The basic functions performed by this system are the following:

\begin{itemize}
\item \textit{Perceive the environment.} The agent interprets the computer image, recognizing the text and the areas where it can click or type.
\item \textit{Evaluate the situation}. The agent analyzes the available information using the language model to attempt to generate a response to the requested task. The available information comes from the user's question and the results of the agent's actions.
\item \textit{Decide actions}. If the agent does not have enough information to generate the response, it uses the language model to generate actions . The agent can generate alternative actions to replace those that did not produce the expected results.
\item \textit{Execute actions}. The agent executes the actions corresponding to mouse and keyboard operations (for example, clicking on links or buttons, entering text, and scrolling) following the generated plan.
\end{itemize}

As indicated, the software architecture of this system includes an LLM as its main component, which is built as a neural network using specific techniques for natural language processing. The pre-trained model is adapted (through fine-tuning or prompting) to interpret computer images and select actions\footnote{For example, WebVoyager uses GPT-4V-Act (\url{https://github.com/ddupont808/GPT-4V-Act}), a large multimodal model, GPT-4V, along with the set-of-mark prompting technique \citep{Yang-2023}}.

\subsection{First-generation expert systems}

In the 1980s, Artificial Intelligence achieved significant commercial success thanks to expert systems. These systems allowed AI to be applied to real-world problems in various sectors, demonstrating that it was possible to automate specialized consulting and diagnostic tasks.

The first generation of expert systems was built primarily using symbolic representations of knowledge, such as rule-based systems. For example, Mycin was one of the pioneering expert systems that served as a reference for the development of many subsequent applications. Mycin was designed to assist in infectious disease diagnosis \citep{Buchanan-1984}.

\begin{figure}[htb!]
\begin{center}
\includegraphics[width=0.70\textwidth]{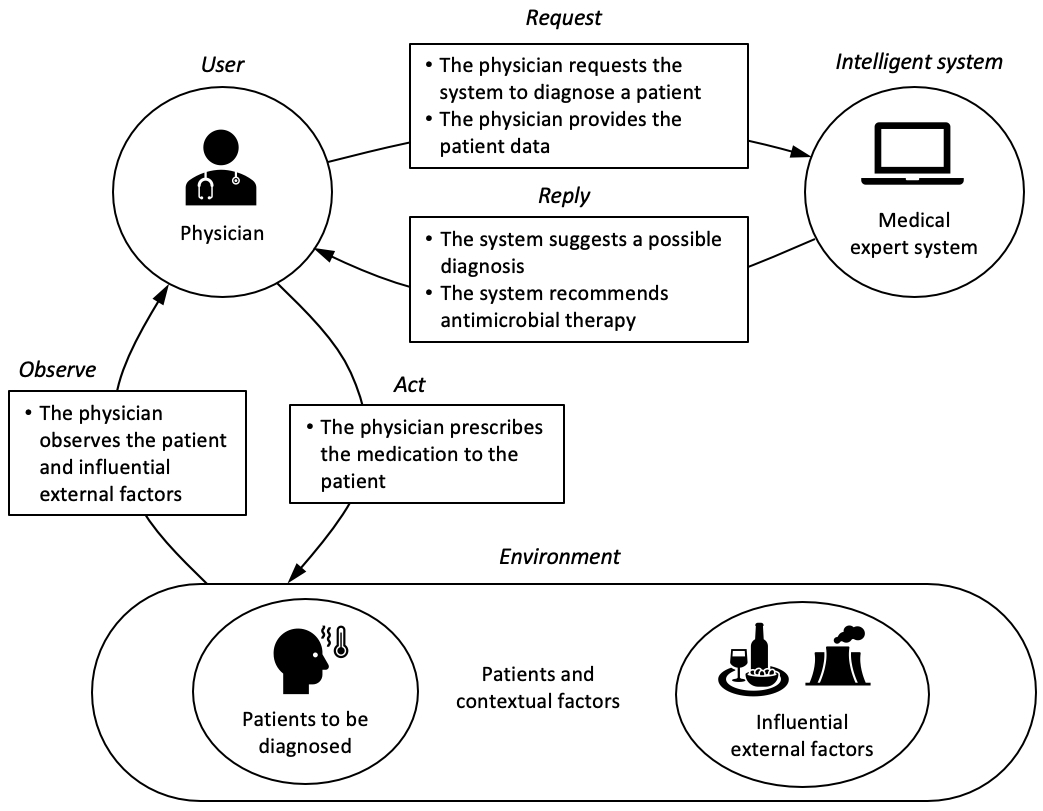}
\caption{Operational context of the infectious disease diagnosis system.}
\label{fig:mycin_example}
\end{center}
\end{figure}

According to the conceptual framework presented in this article, the Mycin system can be analyzed according to the operational context shown in Figure \ref{fig:mycin_example}. In this case, the user is the physician who uses the system as an assistant. The environment includes the patients to be diagnosed and influential external factors (for example, eating and drinking habits, the places where people live and work, etc.).

The properties related to the intelligent behavior of this system are as follows:
\begin{itemize}
\item \textit{Rational acting}. Rational behavior can be evaluated by measuring the degree of success with which the system classifies the organisms that cause infections and the degree of effectiveness of the recommended therapies.
\item \textit{Autonomy}. According to the levels of autonomy described in section \ref{ref:behavioral_properties}, Mycin is classified as level 2 (recommendation system) because it recommends therapies, but it is the physician who prescribes the medication to the patient.
\item \textit{Learning}. Mycin does not learn, as it cannot improve its performance with the repetition of diagnostic tasks.
\item \textit{Explainability}. A key characteristic of Mycin is its ability to provide explanations that justify its conclusions, which is made possible by the symbolic representation of medical knowledge.
\end{itemize}

The basic functions performed by this system are the following:
\begin{itemize}
\item \textit{Perceive the environment.} Information about patients is obtained through communication with the physician. The dialog with the physician takes place in a process of questions and answers with preset sets of potential responses. Mycin asks general questions about the patient (sex, age, etc.) and about cultures of samples taken from parts of the patient's body (blood, saliva, urine, etc.).
\item \textit{Evaluate the situation.} Mycin diagnoses the causes of the symptoms, identifying possible infectious organisms. One of the difficulties of this function is that the system must handle situations in which there is incomplete information about the patient (for example, because culture results from the samples are not available at the time of performing the diagnosis).
\item \textit{Decide actions.} Mycin recommends antimicrobial therapy based on the identified organism and the patient's profile.
\end{itemize}

The software architecture of this system uses a model that represents the diagnostic knowledge of infectious diseases with the following characteristics. The model consists of a knowledge base with about 500 rules that represent the relationships between symptoms and possible infectious organisms. This representation uses an original method of approximate reasoning that employs certainty factors as values in the interval [-1, +1]. The diagnostic process is carried out through a general inference method based on backward chaining and the combination of certainty factors. The rules in the knowledge base were formulated manually with the help of experts in the medical domain.

This example illustrates the application of the conceptual framework presented in this paper to a classic AI system, a first-generation expert system. Systems such as Mycin have been noted for their ability to provide assistance in evaluating the situation (for example, diagnosing with uncertain knowledge) and deciding the actions to be performed, showing understandable explanations that justify their conclusions.

However, these systems are limited mainly by their lack of autonomous operation (as they do not interact directly with the work environment) and their inability to learn from experience. The mechanism they use to communicate with the user is constrained and rigid compared to other forms of communication and data acquisition (for instance, through natural language or sensors), which makes it more difficult to integrate the expert system into day-to-day operations. Furthermore, the knowledge-driven approach used in its construction reduces its applicability to the development of systems with limited-sized models.

\section{Conclusions}  \label{ref:conclusions}
This paper has analyzed the concept of an intelligent system, identifying its general properties and types of components. The paper adopts a descriptive approach as an alternative to the usual classification criteria (the system uses AI methods, acts as a rational agent, or mimics human cognitive abilities), which present problems of circularity, lack of consensus, and difficulty in practical application.

The characterization of the intelligent system is articulated as a conceptual framework divided into four complementary views that homogeneously group its characteristics. This organization makes it possible to distinguish between aspects that are sometimes mixed together when discussing this type of system. The first two views describe the system from an external perspective, addressing respectively its operational context and the properties associated with intelligent behavior. The other two describe its internal organization, distinguishing between the basic functions performed by the system and the types of components that may form part of its software architecture.

The external perspective focuses on the value the system provides within its context of use and on its capacity to effectively help the user achieve their goals, regardless of the specific technology employed. Intelligent behavior is described through properties such as rational action, learning, autonomy, and explainability, which can be present to varying degrees.

The internal perspective complements this view by describing how these capabilities can be realized within the system. To this end, it distinguishes between the basic functions typically performed by intelligent systems (perceiving the environment, assessing the situation, deciding on actions, and executing them) and the types of computational mechanisms that can be used to carry them out.

Taken together, the proposed conceptual framework aims to offer a structured reference for systems engineering students and professionals who need a coherent vocabulary and an organizational scheme for addressing the analysis and construction of intelligent systems. In this sense, the framework should be understood as a conceptual and didactic instrument that complements the primary literature on which it is based.

The applicability of the framework is illustrated through its use in describing systems of diverse natures, both classical and modern, including autonomous robots, decision support systems, agents based on language models, and traditional expert systems. These examples show how different systems can be analyzed and compared using the four views, even though they present different degrees and combinations of the properties associated with intelligent behavior.

Finally, given that artificial intelligence is constantly evolving, with the potential emergence of new computational paradigms of intelligence, the characterization presented in this document, although proposed as stable in general terms, should not be understood as definitive and should be subject to future adaptations to incorporate these advances.

\bibliographystyle{apalike}
\bibliography{references}  

@article{Albus-1991,
  title={Outline for a theory of intelligence},
  author={Albus, James S},
  journal={IEEE transactions on systems, man, and cybernetics},
  volume={21},
  number={3},
  pages={473-509},
  year={1991},
  publisher={IEEE}
}

@article{Albus-1996,
  title={A reference model architecture for design and implementation of intelligent control in large and complex systems},
  author={Albus, James S and Meystel, Alexander},
  journal={International Journal of Intelligent Control and Systems},
  volume={1},
  number={1},
  pages={15--30},
  year={1996}
}

@article{Allen-1999,
  title={Mixed-initiative interaction},
  author={Allen, James E and Guinn, Curry I and Horvtz, Eric},
  journal={IEEE Intelligent Systems and their Applications},
  volume={14},
  number={5},
  pages={14--23},
  year={1999}
}

@article{Anastasi-1992,
  title={What counselors should know about the use and interpretation of psychological tests},
  author={Anastasi, Anne},
  journal={Journal of Counseling \& Development},
  volume={70},
  number={5},
  pages={610--615},
  year={1992},
  publisher={Wiley Online Library}
}

@article{Angele-1998,
  author  = {Angele, Jürgen and Fensel, Dieter and Landes, Dieter and Studer, Rudi},
  title   = {Developing Knowledge-Based Systems with MIKE},
  journal = {Automated Software Engineering},
  volume  = {5},
  pages   = {389--418},
  year    = {1998}
}

@book{Arkin-1998,
  title={Behavior-based robotics},
  author={Ronald C. Arkin},
  year={1998},
  publisher={MIT press}
}

@phdthesis{Bavle-2019,
    title    = {Positioning and Mapping for Aerial Robots using on-board Perception for Autonomous Missions},
    school   = {ETSI Industriales, Universidad Politecnica de Madrid},
    author   = {Hriday Bavle},
    year     = {2019}
}

@article{Beer-2014,
  author  = {Beer, Jenay M. and Fisk, Arthur D. and Rogers, Wendy A.},
  title   = {Toward a Framework for Levels of Robot Autonomy in Human-Robot Interaction},
  journal = {Journal of Human-Robot Interaction},
  year    = {2014},
  volume  = {3},
  number  = {2},
  pages   = {74--99}
}

@article{Bench-Capon-2007,
author = {T.J.M. Bench-Capon and Paul E. Dunne},
title = {Argumentation in artificial intelligence},
journal = {Artificial Intelligence},
volume = {171},
number = {10},
pages = {619-641},
year = {2007}
}

@article{Black-2024,
  author  = {Black, Kevin and Brown, Noah and Driess, Danny and Esmail, Adnan and Equi, Michael and Finn, Chelsea and Fusai, Niccolo and Groom, Lachy and Hausman, Karol and Ichter, Brian and others},
  title   = {$\pi_0$: A Vision-Language-Action Flow Model for General Robot Control},
  journal = {arXiv preprint arXiv:2410.24164},
  year    = {2024}
}

@article{Boden-2008,
  title={Autonomy: What is it?},
  author={Boden, Margaret A},
  journal={Biosystems},
  volume={91},
  number={2},
  pages={305--308},
  year={2008},
  publisher={Citeseer}
}

@article{Bonasso-1997,
  title={Experiences with an architecture for intelligent, reactive agents},
  author={Bonasso, R Peter and Firby, R James and Gat, Erann and Kortenkamp, David and Miller, David P and Slack, Mark G},
  journal={Journal of Experimental \& Theoretical Artificial Intelligence},
  volume={9},
  number={2-3},
  pages={237--256},
  year={1997},
  publisher={Taylor and Francis Ltd}
}

@article{Bonnefon-2016,
  title={The social dilemma of autonomous vehicles},
  author={Bonnefon, Jean-Fran{\c{c}}ois and Shariff, Azim and Rahwan, Iyad},
  journal={Science},
  volume={352},
  number={6293},
  year={2016},
}

@article{Bratman-1988,
  title={Plans and resource-bounded practical reasoning},
  author={Bratman, Michael E and Israel, David J and Pollack, Martha E},
  journal={Computational intelligence},
  volume={4},
  number={3},
  pages={349--355},
  year={1988},
  publisher={Wiley Online Library}
}

@article{Brooks-1986,
  title={A robust layered control system for a mobile robot},
  author={Rodney Brooks},
  journal={IEEE journal on robotics and automation},
  volume={2},
  number={1},
  pages={14--23},
  year={1986},
  publisher={IEEE}
}

@article{Brooks-1991,
  title={Intelligence without representation},
  author={Rodney Brooks},
  journal={Artificial Intelligence},
  volume={47},
  number={1-3},
  pages={139--159},
  year={1991},
  publisher={Elsevier}
}

@book{Brustoloni-1991,
  title={Autonomous agents: Characterization and requirements},
  author={Brustoloni, Jose C},
  year={1991},
  publisher={School of Computer Science, Carnegie Mellon University}
}

@book{Buchanan-1984,
    author  = {B.G. Buchanan and E.H. Shortliffe},
    title = {Rule Based Expert Systems: The MYCIN Experiments of the Stanford Heuristic Programming Project},
    year = {1984},
    publisher = {Addison-Wesley},
    address = {Reading, MA}
}

@incollection{Burgard-2016,
  title={World modeling},
  author={Burgard, Wolfram and Hebert, Martial and Bennewitz, Maren},
  booktitle={Springer handbook of robotics},
  year={2016},
  publisher={Springer}
}

@article{Campbell-2002,
  title={Deep blue},
  author={Campbell, Murray and Hoane Jr, A Joseph and Hsu, Feng-hsiung},
  journal={Artificial intelligence},
  volume={134},
  number={1-2},
  pages={57--83},
  year={2002},
  publisher={Elsevier}
}

@article{Canamero-2001,
  title={Emotions and adaptation in autonomous agents: A design perspective},
  author={Ca{\~n}amero, Lola},
  journal={Cybernetics \& Systems},
  volume={32},
  number={5},
  pages={507--529},
  year={2001},
  publisher={Taylor \& Francis}
}

@inproceedings{Castelfranchi-1994,
  title={Guarantees for autonomy in cognitive agent architecture},
  author={Castelfranchi, Cristiano},
  booktitle={International workshop on agent theories, architectures, and languages},
  pages={56--70},
  year={1994},
  organization={Springer}
}

@inproceedings{Chan-2023,
  title={Harms from increasingly agentic algorithmic systems},
  author={Chan, Alan and Salganik, Rebecca and Markelius, Alva and Pang, Chris and Rajkumar, Nitarshan and Krasheninnikov, Dmitrii and Langosco, Lauro and He, Zhonghao and Duan, Yawen and Carroll, Micah and others},
  booktitle={Proceedings of the 2023 ACM Conference on Fairness, Accountability, and Transparency},
  pages={651--666},
  year={2023}
}

@incollection{Christensen-2016,
  title={Sensing and estimation},
  author={Christensen, Henrik I and Hager, Gregory D},
  booktitle={Springer Handbook of Robotics},
  pages={91--112},
  year={2016},
  publisher={Springer}
}

@book{Conte-1995,
  title={Cognitive and social action},
  author={Conte, Rosaria and Castelfranchi, Cristiano and others},
  year={1995},
  publisher={Garland Science},
  adress={London, UK}
}

@incollection{Coradeschi-2002,
  author = {Coradeschi, S. and Saffiotti, A.},
  title = {Perceptual anchoring: A key concept for plan execution in embedded systems},
  booktitle = {Advances in Plan-Based Control of Robotic Agents},
  publisher = {Springer},
  address = {Berlin, Heidelberg},
  pages = {89-105},
  year = {2002}
}

@book{Crawley-2015,
  title={System architecture: strategy and product development for complex systems},
  author={Crawley, Edward and Cameron, Bruce and Selva, Daniel},
  year={2015},
  publisher={Prentice Hall Press}
}

@incollection{Cuena-1997,
  author    = {Cuena, Jose and Molina, Martin},
  title     = {KSM: An Environment for Design of Structured Knowledge Models},
  booktitle = {Knowledge-Based Systems: Advanced Concepts, Techniques and Applications},
  editor    = {Tzafestas, Spyros G.},
  publisher = {World Scientific Publishing},
  year      = {1997}
}

@incollection{DaSilva-2012,
  author = {Joao Inacio Da Silva Filho and Alexandre Shozo Onuki and Luis Fernando Pompeo Ferrara and Mauricio Conceicao Mario and Jose de Melo Camargo and Dorotea Vilanova Garcia and Marcos Rosa dos Santos and Alexandre Rocco},
  title = {Electric Power System Operation Decision Support by Expert System Built with Paraconsistent Annotated Logic},
  booktitle = {Advances in Expert Systems},
  publisher = {IntechOpen},
  year = {2012}
}

@book{Deacon-1998,
  title={The symbolic species: The co-evolution of language and the brain},
  author={Deacon, Terrence W},
  year={1998},
  publisher={WW Norton \& Company}
}

@inproceedings{Doran-2017,
  title={What Does Explainable AI Really Mean? A New Conceptualization of Perspectives},
  author={Doran, Derek and Schulz, Sarah and Besold, Tarek R.},
  booktitle={Proceedings of the First International Workshop on Comprehensibility and Explanation in AI and ML 2017
co-located with 16th International Conference of the Italian Association for Artificial Intelligence (AI*IA 2017)},
  year={2017},
  url= {https://ceur-ws.org/Vol-2071/CExAIIA_2017_paper_2.pdf},
  address = {Bari, Italy},
  organization={CEUR Workshop Proceedings}
}

@article{Duff-2014,
  title={Maintenance goals in intelligent agents},
  author={Duff, Simon and Thangarajah, John and Harland, James},
  journal={Computational Intelligence},
  volume={30},
  number={1},
  pages={71--114},
  year={2014},
  publisher={Wiley Online Library}
}

@article{Dwivedi-2023,
  title={Explainable AI (XAI): Core ideas, techniques, and solutions},
  author={Dwivedi, Rudresh and Dave, Devam and Naik, Het and Singhal, Smiti and Omer, Rana and Patel, Pankesh and Qian, Bin and Wen, Zhenyu and Shah, Tejal and Morgan, Graham and others},
  journal={ACM Computing Surveys},
  volume={55},
  number={9},
  pages={1--33},
  year={2023},
  publisher={ACM New York, NY}
}

@article{Eisenhardt-1989,
  title={Agency theory: An assessment and review},
  author={Eisenhardt, Kathleen M},
  journal={Academy of management review},
  volume={14},
  number={1},
  pages={57--74},
  year={1989},
  publisher={Academy of Management Briarcliff Manor, NY 10510}
}

@phdthesis{Endsley-1990,
  title={Situation awareness in dynamic human decision making: Theory and measurement},
  author={Endsley, Mica R},
  year={1990},
  school={University of Southern California Los Angeles, CA}
}

@article{Endsley-1999,
  author  = {Endsley, Mica R. and Kaber, David B.},
  title   = {Level of Automation Effects on Performance, Situation Awareness and Workload in a Dynamic Control Task},
  journal = {Ergonomics},
  year    = {1999},
  volume  = {42},
  number  = {3},
  pages   = {462--492}
}

@article{Eriksson-1995,
  author  = {Eriksson, Henrik and Shahar, Yuval and Tu, Samson W. and Puerta, Angel R. and Musen, Mark A.},
  title   = {Task Modeling with Reusable Problem-Solving Methods},
  journal = {Artificial Intelligence},
  volume  = {79},
  pages   = {293--326},
  year    = {1995}
}

@article{Evans-1984,
  title={Heuristic and analytic processes in reasoning},
  author={Evans, Jonathan},
  journal={British Journal of Psychology},
  volume={75},
  number={4},
  pages={451--468},
  year={1984},
  publisher={Wiley Online Library}
}

@article{Evans-2021,
  title={Truthful AI: Developing and governing AI that does not lie},
  author={Evans, Owain and Cotton-Barratt, Owen and Finnveden, Lukas and Bales, Adam and Balwit, Avital and Wills, Peter and Righetti, Luca and Saunders, William},
  journal={arXiv preprint arXiv:2110.06674},
  year={2021}
}

@article{Fernandez-Cortizas-2023,
  author  = {Fernandez-Cortizas, Miguel and Molina, Martin and Arias-Perez, Pedro and Perez-Segui, Rafael and Perez-Saura, David and Campoy, Pascual},
  title   = {Aerostack2: A Software Framework for Developing Multi-Robot Aerial Systems},
  journal = {arXiv preprint arXiv:2303.18237},
  year    = {2023}
}

@inproceedings{Franklin-1996,
  author = {S. Franklin and A. Graesser},
  title = {Is it an Agent, or just a Program?: A Taxonomy for Autonomous Agents},
  booktitle = {International Workshop on Agent Theories, Architectures, and Languages},
  year = {1996}
}

@article{Gabriel-2024,
  title={The ethics of advanced ai assistants},
  author={Gabriel, Iason and Manzini, Arianna and Keeling, Geoff and Hendricks, Lisa Anne and Rieser, Verena and Iqbal, Hasan and Toma{\v{s}}ev, Nenad and Ktena, Ira and Kenton, Zachary and Rodriguez, Mikel and others},
  journal={arXiv preprint arXiv:2404.16244},
  year={2024}
}

@inproceedings{Gat-1999,
  title={Non-linear sequencing and cognizant failure},
  author={Gat, Erann},
  booktitle={AIP Conference Proceedings},
  volume={458},
  pages={660--666},
  year={1999},
  organization={American Institute of Physics}
}

@book{Goertzel-2006,
  title={The hidden pattern: A patternist philosophy of mind},
  author={Goertzel, Ben},
  year={2006},
  publisher={BrownWalker Press}
}

@article{Goodman-2017,
  title={European Union regulations on algorithmic decision-making and a “right to explanation”},
  author={Goodman, Bryce and Flaxman, Seth},
  journal={AI magazine},
  volume={38},
  number={3},
  pages={50--57},
  year={2017}
}

@article{Gottfredson-1997,
  title={Mainstream science on intelligence: An editorial with 52 signatories, history, and bibliography},
  author={Gottfredson, Linda S.},
  journal={Intelligence},
  volume={24},
  number={1},
  pages={13--23},
  year={1997},
  publisher={Ablex Publishing Corporation}
}

@article{Gunning-2019,
  title={XAI—Explainable artificial intelligence},
  author={Gunning, David and Stefik, Mark and Choi, Jaesik and Miller, Timothy and Stumpf, Simone and Yang, Guang-Zhong},
  journal={Science Robotics},
  volume={4},
  number={37},
  year={2019},
  publisher={American Association for the Advancement of Science}
}

@article{Ha-2018,
  title={World models},
  author={Ha, David and Schmidhuber, J{\"u}rgen},
  journal={arXiv preprint arXiv:1803.10122},
  year={2018}
}

@article{HayesRoth-1995,
    author = {Barbara Hayes-Roth},
    title = {An Architecture for Adaptive Intelligent Systems},
    journal = {Artificial Intelligence},
    volume = {72},
    pages = {329-365},
    year =  {1995}
}

@article{Harnad-1990,
  title={The symbol grounding problem},
  author={Harnad, Stevan},
  journal={Physica D: Nonlinear Phenomena},
  volume={42},
  number={1-3},
  year={1990},
  publisher={Elsevier}
}

@article{He-2024,
  title={Webvoyager: Building an end-to-end web agent with large multimodal models},
  author={He, Hongliang and Yao, Wenlin and Ma, Kaixin and Yu, Wenhao and Dai, Yong and Zhang, Hongming and Lan, Zhenzhong and Yu, Dong},
  journal={arXiv preprint arXiv:2401.13919},
  year={2024}
}

@misc{Hoffman-2018,
  author      = {Hoffman, Robert R. and Mueller, Shane T. and Klein, Gary and Litman, Jordan},
  title       = {Metrics for Explainable AI: Challenges and Prospects},
  year        = {2018},
  eprint      = {1812.04608},
  archivePrefix = {arXiv}
}

@article{Huang-2022,
  title={Inner monologue: Embodied reasoning through planning with language models},
  author={Huang, Wenlong and Xia, Fei and Xiao, Ted and Chan, Harris and Liang, Jacky and Florence, Pete and Zeng, Andy and Tompson, Jonathan and Mordatch, Igor and Chebotar, Yevgen and others},
  journal={arXiv preprint arXiv:2207.05608},
  year={2022}
}

@inproceedings{Hughes-2022,
  title={Hydra: a real-time spatial perception system for 3D scene graph construction and optimization},
  author={Hughes, Nathan and Chang, Yun and Carlone, Luca},
  booktitle={Robotics Science and Systems (RSS)},
  year={2022}
}

@techreport{IBM-2005,
  author = {IBM},
  title       = {An Architectural Blueprint for Autonomic Computing},
  institution = {IBM Corporation},
  type        = {White Paper},
  edition     = {3rd},
  year        = {2005},
  month       = {6}
}

@article{Ingrand-2017,
    author = {Felix Ingrand and Malik Ghallab},
    title = {Deliberation for autonomous robots: A survey},
    journal = {Artificial Intelligence},
    volume = {247},
    pages = {10-44},
    year =  {2017}
}

@article{Jo-1997,
  title={Expert system for scheduling in an airline gate allocation},
  author={Jo, Geun-Sik and Jung, Jong-Jin and Yang, Chang-Yoon},
  journal={Expert systems with applications},
  volume={13},
  number={4},
  pages={275--282},
  year={1997},
  publisher={Elsevier}
}

@book{Kahneman-2011,
    author  = {Daniel Kahneman},
    title = {Thinking, fast and slow},
    year = {2011},
    publisher = {Farrar, Straus and Giroux},
    address = {New York}
}

@incollection{Kortenkamp-2016,
  title={Robotic systems architectures and programming},
  author={Kortenkamp, David and Simmons, Reid and Brugali, Davide},
  booktitle={Springer handbook of robotics},
  pages={283--306},
  year={2016},
  publisher={Springer}
}

@article{Kotseruba-2020,
  title={40 years of cognitive architectures: Core cognitive abilities and practical applications},
  author={Kotseruba, Iuliia and Tsotsos, John K},
  journal={Artificial Intelligence Review},
  volume={53},
  number={1},
  pages={17--94},
  year={2020},
  publisher={Springer}
}

@book{Kurzweil-2000,
  title={The age of spiritual machines: When computers exceed human intelligence},
  author={Kurzweil, Ray},
  year={2000},
  publisher={Penguin}
}

@article{Laird-2017,
    author = {J. E. Laird  and C. Lebiere and P. S. Rosenbloom},
    title = {A standard model of the mind: Toward a common computational framework across artificial intelligence, cognitive science, neuroscience, and robotics},
    journal = {AI Magazine},
    volume = {38},
    number = {4},
    pages = {13-26},
    year =  {2017}
}

@article{Langley-2009,
  title={Cognitive architectures: Research issues and challenges},
  author={P. Langley and J. E. Laird and S. Rogers},
  journal={Cognitive Systems Research},
  volume={10},
  number={2},
  pages={141--160},
  year={2009},
  publisher={Elsevier}
}

@article{Lawson-1981,
  title={Command control as a process},
  author={Lawson, Joel},
  journal={IEEE Control Systems Magazine},
  year={1981},
  publisher={IEEE}
}

@article{Legg-2007,
  title={Universal intelligence: A definition of machine intelligence},
  author={Legg, Shane and Hutter, Marcus},
  journal={Minds and machines},
  volume={17},
  number={4},
  pages={391--444},
  year={2007},
  publisher={Springer}
}

@article{Macenski-2022,
  author  = {Macenski, Steve and Foote, Tully and Gerkey, Brian and Lalancette, Chris and Woodall, William},
  title   = {Robot Operating System 2: Design, Architecture, and Uses in the Wild},
  journal = {Science Robotics},
  volume  = {7},
  number  = {66},
  year    = {2022}
}

@article{Maes-1993,
  title={Modeling adaptive autonomous agents},
  author={Maes, Pattie},
  journal={Artificial life},
  volume={1},
  number={1\_2},
  pages={135--162},
  year={1993}
}

@incollection{Marsella-2010,
  title={Computational models of emotion},
  author={Marsella, S. and Gratch, J. and Petta, P.},
  booktitle={A Blueprint for Affective Computing: A Sourcebook and Manual},
  year={2010},
  editor={Scherer, K. R. and Bänziger, T. and Roesch, E. B.},
  publisher={Oxford University Press},
  address={Oxford}
}

@inproceedings{Mayr-2023,
  title={Skiros2: A skill-based robot control platform for ros},
  author={Mayr, Matthias and Rovida, Francesco and Krueger, Volker},
  booktitle={2023 IEEE/RSJ International Conference on Intelligent Robots and Systems (IROS)},
  pages={6273--6280},
  year={2023},
  organization={IEEE}
}

@incollection{Michaud-2016,
  title={Behavior-based systems},
  author={Michaud, Fran{\c{c}}ois and Nicolescu, Monica},
  booktitle={Springer handbook of robotics},
  pages={307--328},
  year={2016},
  publisher={Springer}
}

@article{Miller-2019,
  title   = {Explanation in artificial intelligence: Insights from the social sciences},
  author  = {Miller, Tim},
  journal = {Artificial Intelligence},
  volume  = {267},
  pages   = {1--38},
  year    = {2019},
  publisher = {Elsevier},
  doi     = {10.1016/j.artint.2018.07.007}
}

@book{Mitchell-1997,
    author  = {Tom Mitchell},
    title = {Machine learning},
    year = {1997},
    publisher = {McGraw-Hill}
}

@article{Moerland-2018,
  title={Emotion in reinforcement learning agents and robots: A survey},
  author={Moerland, Thomas M and Broekens, Joost and Jonker, Catholijn M},
  journal={Machine Learning},
  volume={107},
  number={2},
  pages={443--480},
  year={2018}
}

@inproceedings{Molina-2003,
  author = {Martin Molina and Gemma Blasco},
  title = {A multi-agent system for emergency decision support},
  booktitle = {International Conference on Intelligent Data Engineering and Automated Learning},
  year = {2003}
}

@inproceedings{Molina-2005,
  author = {Martin Molina},
  title = {An intelligent assistant for public transport management},
  booktitle = {International Conference on Intelligent Computing},
  year = {2005},
  publisher = {Springer},
  pages = {199-208},
  address = {Berlin, Heidelberg}
}

@book{Murphy-2019,
  title={Introduction to AI robotics},
  author={Robin R. Murphy},
  year={2019},
  publisher={MIT press}
}

@article{Nauta-2023,
  title         = {From Anecdotal Evidence to Quantitative Evaluation Methods: A Systematic Review on Evaluating Explainable AI},
  author        = {Nauta, Meike and Trienes, Jan and Pathak, Shreyasi and Nguyen, Elisa and Peters, Michelle and Schmitt, Yasmin and Schl{\"o}tterer, J{\"o}rg and Van Keulen, Maurice and Seifert, Christin},
  journal       = {ACM Computing Surveys},
  volume        = {55},
  number        = {13s},
  pages         = {1--42},
  year          = {2023},
  publisher     = {ACM New York}
}

@article{Newell-1982,
  title={The knowledge level},
  author={Newell, Allen},
  journal={Artificial intelligence},
  volume={18},
  number={1},
  pages={87--127},
  year={1982},
  publisher={Elsevier}
}

@article{Newell-1976,
  title={Computer science as empirical inquiry: Symbols and search},
  author={Newell, Allen and Simon, Herbert A.},
  journal={Communications of the ACM},
  volume={19},
  number={3},
  year={1976}
}

@book{Newell-1990,
  title={Unified theories of cognition},
  author={Newell, Allen},
  year={1990},
  publisher={Harvard University Press},
  address = {Cambridge, Massachusetts}
}

@article{Niu-2024,
  title={Screenagent: A vision language model-driven computer control agent},
  author={Niu, Runliang and Li, Jindong and Wang, Shiqi and Fu, Yali and Hu, Xiyu and Leng, Xueyuan and Kong, He and Chang, Yi and Wang, Qi},
  journal={arXiv preprint arXiv:2402.07945},
  year={2024}
}

@article{Nwana-1996,
  title={Software agents: An overview},
  author={Nwana, Hyacinth S},
  journal={The knowledge engineering review},
  volume={11},
  number={3},
  pages={205--244},
  year={1996},
  publisher={Cambridge University Press}
}

@article{OCDE-2022,
    author = {OCDE},
    title = {OECD Framework for the Classification of AI systems},
    journal = {OECD Digital Economy Papers},
    volume = {323},
    address = {Paris},
    url = {https://doi.org/10.1787/cb6d9eca-en},
    publisher={OECD Publishing},
    year =  {2022}
}

@book{Pahl-2007,
  title={Engineering design: A systematic approach (3rd edition)},
  author={Gerhard Pahl and Wolfgang Beitz and Jörg Feldhusen and Karl-Heinrich Grote},
  year={2007},
  publisher={Springer}
}

@article{Parisi-2019,
  title={Continual lifelong learning with neural networks: A review},
  author={Parisi, German I and Kemker, Ronald and Part, Jose L and Kanan, Christopher and Wermter, Stefan},
  journal={Neural Networks},
  volume={113},
  pages={54--71},
  year={2019},
  publisher={Elsevier}
}

@misc{Phillips-2020,
  title={Four principles of explainable artificial intelligence},
  author={Phillips, P Jonathon and Hahn, Carina A and Fontana, Peter C and Broniatowski, David A and Przybocki, Mark A},
  howpublished = {National Institute of Standards and Technology (NIST). Internal report 8312},
  year={2020}
}

@book{Picard-2000,
  title={Affective computing},
  author={Picard, Rosalind W},
  year={2000},
  publisher={MIT press}
}

@book{Poole-2010,
  title={Artificial Intelligence: foundations of computational agents},
  author={Poole, David L and Mackworth, Alan K},
  year={2010},
  publisher={Cambridge University Press}
}

@inproceedings{Pronobis-2012,
  title={Large-scale semantic mapping and reasoning with heterogeneous modalities},
  author={Pronobis, Andrzej and Jensfelt, Patric},
  booktitle={2012 IEEE international conference on robotics and automation},
  year={2012},
  organization={IEEE}
}

@inproceedings{Quigley-2009,
  author    = {Quigley, Morgan and Gerkey, Brian and Conley, Ken and Faust, Josh and Foote, Tully and Leibs, Jeremy and Berger, Eric and Wheeler, Rob and Ng, Andrew},
  title     = {{ROS}: An Open-Source Robot Operating System},
  booktitle = {ICRA Workshop on Open Source Software},
  address   = {Kobe, Japan},
  year      = {2009}
}

@article{Rasmussen-1974,
  title={The human data processor as a system component. Bits and pieces of a model},
  author={Rasmussen, Jens},
  year={1974},
  journal={Report No. Riso-M-1722},
  publisher={Danish Atomic Energy Commission},
  adddress={Denmark}
}

@article{Ross-1973,
  title={The economic theory of agency: The principal's problem},
  author={Ross, Stephen A},
  journal={The American economic review},
  volume={63},
  number={2},
  pages={134--139},
  year={1973},
  publisher={American Economic Association}
}

@book{Russell-2014,
    author  = {Stuart Russell and Peter Norvig},
    title = {Artificial intelligence: A modern approach (3rd edition)},
    year = {2014},
    publisher = {Pearson Education}
}

@misc{Sager-2025,
      title={A Comprehensive Survey of Agents for Computer Use: Foundations, Challenges, and Future Directions}, 
      author={Pascal J. Sager and Benjamin Meyer and Peng Yan and Rebekka von Wartburg-Kottler and Layan Etaiwi and Aref Enayati and Gabriel Nobel and Ahmed Abdulkadir and Benjamin F. Grewe and Thilo Stadelmann},
      year={2025},
      eprint={2501.16150},
      archivePrefix={arXiv},
      primaryClass={cs.AI},
      url={https://arxiv.org/abs/2501.16150}, 
}

@article{Sanchez-Lopez-2017,
  title={A multi-layered component-based approach for the development of aerial robotic systems: The Aerostack framework},
  author={Sanchez-Lopez, Jose Luis and Molina, Martin and Bavle, Hriday and Sampedro, Carlos and Suarez Fernandez, Ramon A and Campoy, Pascual},
  journal={Journal of Intelligent \& Robotic Systems},
  volume={88},
  number={2},
  pages={683--709},
  year={2017},
  publisher={Springer}
}

@book{Schreiber-2000,
  author    = {Schreiber, Guus and Akkermans, Hans and Anjewierden, Anjo and de Hoog, Robert and Shadbolt, Nigel and Van de Velde, Walter and Wielinga, Bob},
  title     = {Knowledge Engineering and Management: The CommonKADS Methodology},
  publisher = {MIT Press},
  year      = {2000},
  address   = {Cambridge, MA}
}

@inproceedings{Shakeri-2024,
  title={Formalization of operational domain and operational design domain for automated vehicles},
  author={Shakeri, Ali},
  booktitle={IEEE 24th International Conference on Software Quality, Reliability, and Security Companion (QRS-C)},
  year={2024}
}

@article{Shapiro-2005,
  title={Agency theory},
  author={Shapiro, Susan P},
  journal={Annual Review of Sociology},
  volume={31},
  pages={263--284},
  year={2005},
  publisher={Annual Reviews}
}

@article{Shavit-2023,
  title={Practices for governing agentic AI systems},
  author={Shavit, Yonadav and Agarwal, Sandhini and Brundage, Miles and Adler, Steven and O’Keefe, Cullen and Campbell, Rosie and Lee, Teddy and Mishkin, Pamela and Eloundou, Tyna and Hickey, Alan and others},
  journal={Research Paper, OpenAI},
  year={2023}
}

@article{Sheridan-1978,
  author  = {Sheridan, Thomas B. and Verplank, William L.},
  title   = {Human and Computer Control of Undersea Teleoperators},
  journal = {MIT Man-Machine Systems Laboratory Technical Report},
  year    = {1978}
}

@article{Simon-1955,
  title={A behavioral model of rational choice},
  author={Simon, Herbert A},
  journal = {The Quarterly Journal of Economics},
  pages={99--118},
  year={1955},
  volume = {69},
  publisher = {Oxford University Press}
}

@article{Turpin-2023,
  title={Language models don't always say what they think: Unfaithful explanations in chain-of-thought prompting},
  author={Turpin, Miles and Michael, Julian and Perez, Ethan and Bowman, Samuel},
  journal={Advances in Neural Information Processing Systems},
  volume={36},
  pages={74952--74965},
  year={2023}
}

@book{Vernon-2014,
  title={Artificial cognitive systems: A primer},
  author={Vernon, David},
  year={2014},
  publisher={MIT Press}
}

@incollection{Vernon-2022,
    author = {Vernon, David},
    title = {Cognitive Architectures},
    booktitle = {Cognitive Robotics},
    publisher = {The MIT Press},
    editor = {Angelo Cangelosi and Minoru Asada},
    year = {2022}
}

@misc{Wang-2024,
      title={Mobile-Agent: Autonomous Multi-Modal Mobile Device Agent with Visual Perception}, 
      author={Junyang Wang and Haiyang Xu and Jiabo Ye and Ming Yan and Weizhou Shen and Ji Zhang and Fei Huang and Jitao Sang},
      year={2024},
      eprint={2401.16158},
      archivePrefix={arXiv},
      primaryClass={cs.CL},
      url={https://arxiv.org/abs/2401.16158} 
}

@article{Wang-2024b,
  title={Mobile-agent-v2: Mobile device operation assistant with effective navigation via multi-agent collaboration},
  author={Wang, Junyang and Xu, Haiyang and Jia, Haitao and Zhang, Xi and Yan, Ming and Shen, Weizhou and Zhang, Ji and Huang, Fei and Sang, Jitao},
  journal={Advances in Neural Information Processing Systems},
  volume={37},
  pages={2686--2710},
  year={2024}
}

@article{Wooldridge-1995,
    author = {M. J. Wooldridge and N. R. Jennings},
    title = {Intelligent agents: Theory and practice},
    journal = {The knowledge engineering review},
    volume = {10},
    number = {2},
    pages = {115-152},
    year =  {1995}
}

@incollection{Wooldridge-1999,
  title={Intelligent agents},
  author={M. J. Wooldridge},
  booktitle={Multiagent systems: A modern approach to distributed artificial intelligence},
  year={1999},
  editor = {Weiss, Gerhard},
  publisher={MIT press}
}

@book{Wooldridge-2009,
    title={An introduction to multiagent systems (Second ed)},
    author={M. J. Wooldridge},
    year = {2009},
    publisher = {John Wiley and Sons}
}

@article{Xi-2025,
  title={The rise and potential of large language model based agents: A survey},
  author={Xi, Zhiheng and Chen, Wenxiang and Guo, Xin and He, Wei and Ding, Yiwen and Hong, Boyang and Zhang, Ming and Wang, Junzhe and Jin, Senjie and Zhou, Enyu and others},
  journal={Science China Information Sciences},
  volume={68},
  number={2},
  pages={121101},
  year={2025},
  publisher={Springer}
}

@article{Yang-2023,
  title={Set-of-mark prompting unleashes extraordinary visual grounding in gpt-4v},
  author={Yang, Jianwei and Zhang, Hao and Li, Feng and Zou, Xueyan and Li, Chunyuan and Gao, Jianfeng},
  journal={arXiv preprint arXiv:2310.11441},
  year={2023}
}

@inproceedings{Yao-2023,
  title     = {{ReAct}: Synergizing Reasoning and Acting in Language Models},
  author    = {Shunyu Yao and Jeffrey Zhao and Dian Yu and Nan Du and Izhak Shafran and Karthik Narasimhan and Yuan Cao},
  booktitle = {Proceedings of the 11th International Conference on Learning Representations (ICLR)},
  year      = {2023},
  address   = {Kigali, Rwanda}
}






\end{document}